\documentclass[a4paper,10pt]{article}
\usepackage{amsmath,amssymb,amsthm}
\usepackage{cite}
\usepackage{indentfirst}
\usepackage{graphicx}

\def\eps{\varepsilon}
\def\psie{\psi^{\eps}}
\def\cpsie{\overline{\psi}^{\eps}}

\def\We{W^{\eps}}
\def\tWe{{\widetilde W}^{\eps}}
\def\Sh{\mathcal{S}}
\def\R{\mathbb{R}}

\def\N{\mathbb{N}}
\def\Hop{\widehat{H}}
\def\ue{u^{\eps}}

\def\Ee{E^{\eps}}

\def\Le{\mathcal{L}^{\eps}}
\def\Me{\mathcal{M}^{\eps}}
\def\Lc{\mathcal{L}_{c}}
\def\LH{\mathcal{L}_{h}}
\def\MH{\mathcal{M}_{h}}

\def\Phinm{\Phi^{\eps}_{nm}}
\def\Psinm{\Psi^{\eps}_{nm}}
\def\Znm{Z_{nm}}
\def\tZnm{{\widetilde Z}_{nm}}

\def\cx{\xi}
\def\ck{\eta}
\def\pk{\frac{\partial}{\partial{\eta}}}
\def\pkj{\frac{\partial^{(2j+1)}}{\partial{\eta^{2j+1}}}}
\def\pkkj{\frac{\partial^{(2j)}}{\partial{\eta^{2j}}}}
\def\pkk{\frac{\partial^{(2)}}{\partial{\eta^{2}}}}
\def\px{\frac{\partial}{\partial{\xi}}}
\def\pkkk{\frac{\partial^{(3)}}{\partial{\eta^3}}}
\def\pklj{\frac{\partial^{(2\lambda_j+1)}}{\partial{\eta^{2\lambda_j+1}}}}
\def\pkl{\frac{\partial^{(2\lambda_1+1)}}{\partial{\eta^{2\lambda_1+1}}}}
\def\B{\mathcal{B}}
\def\Anoe{A_{0,n}^{\eps}}
\def\Amoe{A_{0,m}^{\eps}}

\newtheorem{lemma}{Lemma}
\newtheorem{thm}{Theorem}
\newtheorem{prop}{Proposition}

\begin{document}

\title{Perturbation solutions of the \\ semiclassical Wigner equation}

\author{E.K. Kalligiannaki\footnote{E-mail:vageliwka@gmail.com} 
 \& G.N. Makrakis \footnote{E-mail:makrakg@iacm.forth.gr} 
\footnote{Also: Institute of Applied and Computational Mathematics, 
FORTH, 71110 Heraklion, Crete, Greece}\\
Department of Applied Mathematics, \\
University of Crete,  71409  Heraklion, Crete, \\
Greece
}
                              
\maketitle

\begin{abstract}
We present a perturbation analysis of the semiclassical Wigner equation  which is based on the interplay between configuration and phase spaces via Wigner transform. We employ the so-called harmonic approximation of the Schr\"odinger eigenfunctions for single-well potentials in configuration space, to construct an asymptotic expansion of the solution of the Wigner equation. This expansion  is a  perturbation of the Wigner function of a harmonic oscillator but it is not a genuine semiclassical expansion because the correctors depend on the semiclassical parameter.  However, it suggests the selection of a novel  ansatz for the solution of the Wigner equation, which leads to an efficient  regular perturbation scheme in phase space. The validity of the approximation is proved for particular classes of initial data.  The proposed ansatz is applied  for computing the energy density  of a a quartic oscillator on caustics (focal points). The results are compared with those derived from the so-called classical approximation whose  principal term is the solution of  Liouville equation with the same initial data. It turns out that the results are in good approximation when the coupling constant of the anharmonic potential has certain dependence on the semiclassical parameter.
\end{abstract}

\noindent
\section{ Introduction}
\setcounter{equation}{0}
\renewcommand{\theequation}{1.\arabic{equation}}

\noindent
\paragraph{WKB solutions and geometrical optics.}

We consider the oscillatory initial value problem

\begin{equation}\label{Schro}
\left\{
\begin{array}{l}
i\eps \frac{{\partial \psie }}{{\partial t}}= \Hop \psie(x,t),\ \ x\in
\R,\ t>0 \\
\psie(x,0)=\psie_0(x) \ , \ \  \eps<<1 \ ,
\end{array}
\right.
\end{equation}
where $\Hop = -\frac{\eps^2}{2}\partial^{2}_{x} + V(x)$ is the usual quantum mechanical 
Schr\"odinger operator. High frequency waves that satisfy (\ref{Schro})  with oscillatory initial data
\begin{equation}\label{wkb0}
\psie_0(x)=A_0(x)\exp(i S_0(x)/\eps) \ ,
\end{equation}
have been traditionally studied by geometrical optics. This method departs from the construction of WKB  approximate solutions  in the form (\cite{BABU}, \cite{BLP}, \cite{KO})
\begin{equation}\label{wkb}
\psie (x,t) = A(x,t)\exp(i S(x,t)/\eps) \ ,
\end{equation}
where the phase function $S(x,t)$ and the amplitude $A(x,t)$ are  usually assumed to be real-valued functions, although extensions of the method for complex-valued phases have been also developed.

Substituting (\ref{wkb}) into (\ref{Schro}), and retaining terms of order $O(\eps)$ and $O(1)$ with respect to  $\eps$,  we obtain the Hamilton-Jacobi equation for the phase  

\begin{equation}\label{eikonal}
\partial_{t}S + (\partial_{x}S)^{2}/2 +V =0 \ , \ \ \ \ S(x, t=0)= S_{0}(x) \ ,
\end{equation}
and the transport equation for the amplitude 

\begin{equation}\label{transport}
\partial_{t}(A^2) +\partial_{x}\left(A^2\partial_{x}S\right)  =0
\ , \ \ \ \ A(x,t=0)=A_{0}(x) \ .
\end{equation}

The system (\ref{eikonal})-(\ref{transport})  is integrated byreduction to a system of ordinary differential equations along bicharacteristics as follows. We  define  the Hamiltonian function
$$
H(x,k)= k^{2}/2 + V(x) \ ,
$$
and construct the characteristics  $(\tilde{x}(t;q),
\tilde{k}(t;q))$, as the trajectories of the Hamiltonian system

\begin{eqnarray}\label{hamsy}
\frac{d\tilde{x}}{dt}=H_{k}(\tilde{x}, \tilde{k})=\tilde{k} \ , \
\ \ \ \ \frac{d\tilde{k}}{dt}=-H_{x}(\tilde{x}, \tilde{k})=V'(\tilde{x})
\end{eqnarray}
with initial conditions
$$
\tilde{x}(t=0;q)= q \ , \ \ \ \ \ \ \tilde{k}(t=0;q)= S'_{0}(q) \
$$
The projection of the characteristics $\tilde{x}=\tilde{x}(t;q)$ on the configuration space are the physical rays of geometrical optics. Then,  the phase function $S$ is obtained by integration of the ordinary differential equation
$$
\frac{dS}{dt}= \partial_{t}S + \partial_{x}S\frac{dx}{dt}=-H(\tilde{x}(t;q), \tilde{k}(t;q)) + \tilde{k}^{2}(t;q)/2=\tilde{k}^{2}(t;q) -V(\tilde{x}(t;q)) \ ,
$$
with the initial condition
$$
S(\tilde{x}(t=0;q))=S_{0}(q) \ .
$$

Furthermore, the amplitude $A$ is calculated  by applying the divergence theorem in a ray tube, and is given by
$$
A(\tilde{x}(t;q),t)= \frac{A_{0}(q)}{\sqrt{J(t;q)}} \ ,
$$
where
$$
J(t;q)= \frac{\partial \tilde{x}(t;q)}{\partial q}
$$
is the Jacobian of the ray transformation $q \mapsto\tilde{x}(t;q)$.

The nonlinear Hamilton-Jacob equation has, in general, multivalued solutions. This means that   singularities may be formatted  in finite time, when the Jacobian of the ray transformation vanishes. The set $\mathcal {C}=\{x= \tilde{x}(t;q) :J(t;q)=0 \} $ of  these singularities is the caustic of the ray field. On the caustics  the amplitude  becomes infinite. Therefore,  near such singularities the WKB method fails to predict the correct wave field.  in the sense that the method cannot describe the correct scales. These scales are predicted either from analytical solutions of model problems, or from uniform asymptotic expansions for the solution of the Schr\"{o}dinger equation. Such solutions  show that the amplitude of the intensity of the wave field increases with the frequency $\eps^{-1}$, but, for any fixed large frequency, it remains bounded with respect to space variables on the caustics.

 Assuming that the multivalued function $S$ is  known away from caustics, and using boundary layer techniques and matched asymptotic expansions \cite{BABU}, \cite{BK}, it has been   possible to constructed uniform asymptotic expansions near simple caustics. However these analytical techniques are very  complicated  since  the matching procedure depends on the form of the particular caustic.

A different group of  uniform methods valid near caustics, is based on integral representations of the solutions in phase-space. The basic methods in this category are  the Maslov's canonical operator\cite{MF}. \cite{MSS}, and the Lagrangian integrals (Kravtsov-Ludwig method \cite{KRA}, \cite{Lu},  \cite{KO}), which can be considered as  special cases of  Fourier integral operators \cite{Du}, \cite{TRE}.

All the  above described techniques  assume an ansatz for the solution in the configuration space, which for the final determination requires the knowledge of the multivalued phase functions, or, geometrically, of the Lagrangian manifold generated by the bicharacteristics of the Hamiltonian system  in phase space.

 An alternative approach  is based on the use of the Wigner transform. This is a function defined on phase space as the Fourier transform of the two-point correlation of the wave function. It was  introduced by Wigner \cite{Wig} for specific purposes  in quantum thermodynamics, and recently it has been successfully used in semiclassical analysis for the   reformulation of wave equations  as non local equations in phase space, and the study of  of homogenization problems in high-frequency waves \cite{GMMP}. 

\noindent 
\paragraph{ The Wigner equation.}

The Wigner transform of  a function $f \in L^2$,  is defined as
\begin{equation}\label{wignerd}
\We[f](x,k)=\frac{1}{2\pi\eps}\int_\R e^{-\frac{i}{\eps}k\xi} f(x+\frac{\xi}{2})\overline{f}(x-\frac{\xi}{2}) d\xi
\end{equation}
and for a pair of  functions $ f, g\in L^2(\R)$, the (cross) Wigner transform  is defined as
 \begin{equation}\label{wignerfg}
\We[f,g](x,k)=\frac{1}{2\pi\eps}\int_\R e^{-\frac{i}{\eps}k\xi} f(x+\frac{\xi}{2})\overline{g}(x-\frac{\xi}{2}) d\xi \ .
\end{equation}

By its definition  $\We[f](x,k)$   is  a real , square integrable,   but  not necessarily non-negative function in phase space. For this reason it is not a pure probability distribution function,  but it is exactly this property that makes the Wigner function a powerful tool in the study of wave and quantum interference phenomena. 

Some of the most important relations of Wigner functions 
\begin{equation}\label{wignfunc}
\We(x,k):=\We[\psie](x,k)
\end{equation}
with the wave function $\psie(x)$ (this in general depends also on time $t$), that are useful for the computation of physical quantities  both in classical wave propagation and in quantum mechanics, are the following 

 \medskip
 \noindent
 1) The integral of $\We$ with respect to the momentum $k$ gives   the energy density    
 \begin{equation}\label{ampl}
 \eta^{\eps}(x):= |\psie(x)|^{2}=\int_{\R} \We(x,k)dk \ , 
 \end{equation}
 while  the first moment  with respect to the momentum  gives the energy flux, 
\begin{equation}
\mathcal{F^{\eps}}(x) := \frac{\eps}{2i}\Bigl(\cpsie
\partial_{x}\psie -\psie \partial_{x} \cpsie \Bigr)=\int_{\R} k \We(x,k)dk
\end{equation}
 and its integral over the whole phase space, gives the total energy 
 \begin{equation}
 \int_{\R}\int_{\R}dxdk \We(x,k)=\| \psie\|_{L^2} \ .
  \end{equation}

\medskip
 \noindent
 2) The Wigner  transform of a  WKB function   
 $$ \psie (x) = A(x)\exp(i S(x)/\eps) \ ,$$  
 considered as a generalized function,  as $\eps \rightarrow 0$, has the weak limit
\begin{equation}\label{wkblimit}
\We[\psie](x,k)\rightharpoonup |A(x)|^{2}\delta\bigl(k- S'(x)\bigr) \ ,
\end{equation}
where $\delta$ denotes the Dirac distribution \cite{LP}, \cite{PR}.

The evolution equation of the Wigner function $\We(x,k,t):=\We[\psie](x,k,t)$   corresponding to the solution $\psie(x,t)$ of (\ref{Schro}) has the form\footnote{This equation  is referred in the context of classical wave propagation  as the Wigner equation \cite{PR} and in quantum mechanics as the quantum Liouville equation \cite{MA}}

\begin{equation}\label{tql}
\left\{
\begin{array}{l}
\frac{\partial}{\partial t}\We(x,k,t)+\Le\We(x,k,t)=0, (x,k)\in\R^2, t>0\\
\We(x,k,t)|_{t=0}=\We_0(x,k)
\end{array}
\right.
\end{equation}
where $\We_0(x,k)$ is the Wigner transform of the initial data   $\psie_0(x)$. and $\Le$ is the  pseudo differential operator defined by (\cite{BA}, \cite{FAI})

\begin{equation}\label{Les}
\Le \cdot := -\frac{2}{\eps}\sin\Bigl(\frac{\eps}{2} \Lambda \Bigr)H(x,k)  \cdot
\end{equation}
For a general Hamiltonian $H(x,k)$ the operator $\Lambda$ is the commutator 
$$
\Lambda:=\frac{\partial}{\partial
x_H}\frac{\partial}{\partial k}-\frac{\partial}{\partial
x}\frac{\partial}{\partial k_H} \ ,
$$
and $\sin\Bigl(\frac{\eps}{2} \Lambda \Bigr)$ acts on the product of two functions $f(x,k)$ and $g(x,k)$ by the formula
\begin{align*}
 &\sin\Bigl(\frac{\eps}{2} \Lambda \Bigr)f(x,k) g(x,k)=\\
&=\frac{1}{2i\pi^2\eps^2}\int dk'dk''dx'dx''\left[f(x',k')g(x'',k'')-g(x',k')f(x'',k'')\right]\\
&\times \exp\left(-\frac{2i}{\eps}\left(k(x'-x'')+
k'(x''-x)+k''(x-x')\right)\right) \ .
\end{align*}

We also introduce for later use (see eq ($\ref{eig2}$) in Section 2) the  pseudo-differential operator 
\begin{equation}\label{Mes} 
 \Me \cdot :=\cos\Bigl(\frac{\eps}{2} \Lambda \Bigr)H(x,k)\cdot 
\end{equation} 
 which is   known in physics' literature as Baker's cosine bracket, by  (\cite{CFZ}, \cite{FAI}, \cite{TH}), and $\cos\Bigl(\frac{\eps}{2} \Lambda \Bigr)$ acts on the product of two functions $f(x,k)$ and $g(x,k)$ as follows   
\begin{align*}
&\cos\Bigl(\frac{\eps}{2} \Lambda \Bigr)f(x,k) g(x,k)=\\
&=\frac{1}{2\pi^2\eps^2}\int dk'dk''dx'dx''\left[f(x',k')g(x'',k'')+g(x',k')f(x'',k'')\right]\\
&\times\exp\left(-\frac{2i}{\eps}\left(k(x'-x'')+
k'(x''-x)+k''(x-x')\right)\right) \ .
\end{align*}

 In the particular case of the usual quantum mechanical Hamiltonian $H(x,k)=k^2/2 + V(x)$, 
 by using (\ref{Les}), the operator  $\Le$ may also be written in the more standard  form
\begin{equation}
\Le=k\partial_x-\Theta^{\eps}[V]
\end{equation}
where the operator $\Theta^{\eps}[V]$ is expresses as the convolution
\begin{equation}
\Theta^{\eps}[V]\We(x,k,t):=Z^{\eps}(x,k)*_k\We(x,k,t) \ ,
\end{equation}\label{convtheta}
of the Wigner function with the kernel
\begin{equation}\label{zeta}
Z^{\eps}(x,k)=\frac{1}{i\eps}\frac{1}{2\pi}\int_{\R}e^{-iky}\left(V(x+\frac{\eps}{2}y)-
V(x-\frac{\eps}{2}y)\right)dy \ .
\end{equation}
This is a non-local operator which the action of the potential on the evolution of Wigner function. 

From and(\ref{convtheta})  (\ref{zeta}) we can write the action of $\Theta^{\eps}[V]$ as a pseudo-differential operator
\begin{equation}\label{theta}
\Theta^{\eps}[V]\We(x,k,t)=\frac{i}{2\pi\eps}\int_{\R}\int_{\R}e^{i(k-\xi)y}\We(x,\xi,t)\left[V(x+\frac{\eps}{2}y)-V(x-\frac{\eps}{2}y)\right]dyd\xi \ .
\end{equation}

Now we observe that if the potential function $V$ is smooth enough, using its Taylor expansion we may rewrite quantum Liouville or Wigner operator $\Le$ in the form of an infinite order differential operator 

\begin{equation}\label{Le}
 \Le= k\frac{\partial}{\partial x}-V'(x)\frac{\partial}{\partial k}-
\sum_{j=1}^{\infty}\eps^{2j}\left(\frac{i}{2}\right)^{2j}\frac{V^{(2j+1)}(x)}{(2j+1)!}\frac{\partial^{(2j+1)}}{\partial k^{2j+1}} \ .
\end{equation}
Thus the Wigner equation (\ref{tql}) is written in  the form   of an infinite order ingular perturbation  equation
\begin{equation}\label{wignereql}
\partial_{t}\We + k\partial_{x}\We- V'(x) \partial_{k}\We = \sum_{m=1}^{\infty} \alpha_m
\eps^{2m} V^{(2m+1)} (x)
\partial_{k}^{2m+1} \We(x,k,t)  \ ,
\end{equation}
where
 $\alpha_m = (-1)^m/2^{2m} (2m+1)!,\  \ m=0,1, \dots$, and
 $V^{(2m+1)} (x) =d^{2m+1} V(x)/dx^{2m+1}$.

The form  (\ref{wignereql}) of the Wigner equation shows that that this equation is  a combination of the  classical transport  (Liouville) operator in the left hand side, with  a dispersion operator of infinite order in he right hand side.  Roughly speaking, this combination  suggests that  the  phase space evolution results from the  interaction between the classical transport of the Lagrangian manifold generated by the Hamiltonian and a non-local dispersion of energy from the manifold into the whole phase space. This picture is consistent with the fact that in the classical limit $\eps =0$ the dispersion mechanism disappears. Then,  the solution of the Wigner equation converges weakly to the so called Wigner measure \cite{LP}, which satisfies the Liouville equation of classical mechanics. This solution is an always well-defined semiclassical measure, and in the absence of caustics it completely retrieves the results of the WKB method

However,  it has been  shown  in  \cite{FM1}, \cite{FM2} that in the case of multi-phase optics and caustic formation, the limit Wigner measure, although still well-defined as semiclassical measure on phase space, is not the appropriate tool for the computation of energy densities at a fixed point of configuration space, because; (a) it cannot be expressed as a distribution with respect to the momentum for a fixed space-time point, and thus it cannot be used to compute the amplitude of the wavefunction, on caustics, and (b) it is unable to "recognise" the correct scales of the wavefield  near caustics. It has been also shown in \cite{FM1} that approximate Airy-type solutions of the Wigner equation can,  produce reasonable solutions for multiphase problems, and at least for simple caustics,

Therefore the study of asymptotic solutions of the Wigner equation  for small $\eps$ is promising  for understanding the structure  solutions, and for computing energy densities, in multiphase geometrical optics through integration of  the Wigner function.
 
Several asymptotic solutions of the Wigner equation have been proposed in the recent past. Steinr\"uck \cite{ST} and Pulvirenti \cite{PU},  have constructed distributional asymptotic expansions near the solution of the classical Liouville equation, by expanding the initial data in a distributional series with respect to the small parameter. However, Heller  \cite{EH} has noted  that such expansions are not physically appropriate for  studying the evolution of singular initial conditions, and he has proposed a different expansion where the first order term is the solution of a classical Liouville equation with an effective potential. The use of modified characteristics and effective potentials  aims to include indirectly some quantum phenomena  and it is a popular technique in physics and quantum chemistry  for the treatment of the quantum Liouville equations (see, e.g.,  the review paper \cite{HWL}). It has led to reasonable numerical results, and, somehow, it can be used as an alternative of  quantum hydrodynamics (Bohm equations)  and of the technique of Gaussian beams. In the same direction, Narcowich \cite{FN1} proposed a different expansion near the classical Liouville equation without expanding the  initial data with respect to the semiclassical parameter, which allows him to  avoid  the distributional expansions.
In a rather different direction,  the uniform Airy type asymptotic approximation of the Wigner function that  was proposed by  Berry \cite{BE}, has been used in the works of Filippas \& Makrakis \cite{FM1}, \cite{FM2}, to contract novel  asymptotic solution of the Wigner equation  in the presence of simple caustics.

\noindent
\paragraph{Outline of the paper.}

This paper aims to  the understanding of asymptotic solutions of Wigner equation, by adopting a new strategy for the construction of asymptotic expansions, that exploits the interplay between configuration and phase spaces via Wigner teansform. Our approach has been  motivated by the general  idea of using  spectral expansions in the construction of high-frequency solutions, which  has been developed in \cite{BLP}, Ch. 4, for Schr\"odinger equations. Eigenfunction expansions can be considered as `exact'' solutions of the Cauchy problem ($\ref{Schro}$), which in contrary to the WKB solutions,  do not face caustic problems. When transferred to the phase space by Wigner transform they give corresponding expansions which are "exact" solutions of the Wigner equation. 

In this respect, our strategy is the following:

\noindent
(1) We construct the Moyal eigenfunctions of the Wigner equation (Section 2.1) and their asymptotic expansions  in terms of the Moyal functions  of a harmonic oscillator, which arises from the so-called harmonic approximation of Schr\"odinger eigenfunctions (Section 2.2). We assume a single well potential $V(x)$,   so that  the Schr\"odinger spectrum be  purely discrete, and spectral information for the Wigner equation be also available.

\noindent
(2) We transform the eigenfunction expansion of the Schr\"odinger equation to the phase space, and we construct  the solution of the Wigner equation as a series of phase-space Moyal eigenfunctions (Section 3.1). Then, we construct a {\it formal} asymptotic expansion (harmonic expansion) for the solution of Wigner equation using the expansions of Moyal eigenfunction derived in the first step, and finally

\noindent
(3) We  propose an {\it ansatz} for the solution of the Wigner equation and we develop   a regular-perturbation scheme in phase space, for computing  the sought for asymptotic expansion for the solution of the Wigner equation (Section 3.2).

In Section 4 we present the so-called classical expansion, where the solution of the Wigner equation is expressed as a perturbation of the solution of classical Liouville equation.

As an application,   in Section  5 we apply the proposed scheme for a quartic (anharmonic)  oscillator, and we compute the wave amplitude at the beaks of the cusps generated by the oscillator through integration of the  approximate Wigner functions with respect to the momentum. The predictions of the harmonic and the classical expansions agree at the singular points provided that  a certain relation between the small semiclassical parameter and the coupling constant of the potential holds.

\noindent
\section{Aproximmation of Moyal eigenfunctions}
\setcounter{equation}{0}
\renewcommand{\theequation}{2.\arabic{equation}}

\noindent
\subsection{The Moyal eigenfunctions  in phase space}

It is known that, in principle, the spectrum of the quantum Liouville operator  can be  determined from the   spectrum of  the corresponding Schr\"odinger  operator $\Hop$ (see, e.g.  \cite{MA},  \cite{SP}), and, in general,  someone anticipates the formula
$$\sigma(\Le)=\{\frac{i}{\eps}(E-E'), \ \ E, \ E' \in \sigma(\Hop)\} \  , $$
 to hold.
In fact, this  relation  holds   for the discrete spectrum 
$$\sigma_p(\Le)=\{\frac{i}{\eps}(\Ee_n-\Ee_m), \ \ \Ee_n, \Ee_m \in \sigma_p(\Hop)\} \  \ .$$
A similar formula  holds for the point spectrum of the cosine bracket operator $\Me$ (eq ($\ref{Mes}$) below), that is 
$$\sigma_p(\Me)=\{\frac{1}{2}(\Ee_n+\Ee_m), \ \ \Ee_n, \Ee_m \in \sigma_p(\Hop)\} \ .$$ 

However, these formulae  are not in general true for the absolutely and singular continuous spectrum. These spectral questions  have been studied first  by Spohn  \cite{SP}, and later by Antoniou et al  \cite{ASS}, who have proved  the negative result
 $$\sigma_{sc,ac}(\Le)\neq  \{\frac{i}{\eps}(E-E'), \ \ E, \ E' \in \sigma_{sc,ac}(\Hop)\} \ , $$
 where $\sigma_{sc,ac}$ denote the singular and absolutely continuous  spectrum respectively.

In order to avoid the complications arising from the continuous spectrum (a;though this pertains to the most interesting cases of scattering problems),  we consider operators $\Hop$ with purely discrete spectrum $\sigma(\Hop)=\sigma_p(\Hop)$, therefore $\sigma(\Le)=\sigma_p(\Le) $ and $\sigma(\Me)=\sigma_p(\Me).$
When the   potential    $V(x)\in L^1_{loc}(\R)$ is bounded below and $  \lim_{|x|\rightarrow \infty}V(x)=\infty $  it is known that $\Hop$ has purely discrete spectrum ( \cite{RSIV}, \cite{BS}), and therefore the operators $\Le, \Me$ have also purely discrete spectrum. We denote by  $\Ee_n$ and  $\ue_{n}(x)$  the eigenvalues and   the eigenfunctions of $\Hop$, that satisfy
$\Hop \ue_{n}(x)=\Ee_n\ue_{n}(x) \ , \ \ n=1,2, \dots \ .$ It is  known that $\ue_{n}$ form a complete orthonormal basis in $L^2(\R)$ .

The Moyal  eigenfunctions,  were introduced  by Moyal \cite{MO}, for the purposes of a concrete statistical study of quantum mechanics, and they are
  defined as  the cross-Wigner   transform (\ref{wignerfg}) of the Schr\"odinger 
eigenfunctions $\ue_{n} \ , \ue_{m}  \ ,\  \ \  n,m=0,1,2,\dots$. 

 \begin{equation}\label{phinm}
\Phinm(x,k):=\We[\ue_{n},\ue_{m}](x,k)= \frac{1}{\pi\eps}\int_{\mathbf{R}}e^{-i\frac{2k}{\eps}\sigma}
\ue_{n}(x+\sigma)\overline{\ue_{m}}(x-\sigma)d\sigma \ .
\end{equation}

For these functions the following theorem holds \cite{MO}, \cite{TH}.
\begin{thm} 
Let $\Hop$    has  purely  discrete spectrum $\{\Ee_n\}_{n=0,1,2,\dots}$  with  complete orthonormal 
system of eigenfunctions $\{\ue_{n}(x)\}_{n=0,1,\dots}$ in $   L^2(\R) $. 
Then, the functions $\{\Phi^{\eps}_{nm}\}_{ n,m=0,1,\dots} $  form a complete orthonormal basis in $L^2(\R_{xk}^2)$, and they are common eigenfunctions of the operators  
$\Le$ and  $\Me$
 with eigenvalues $\lambda_{nm}=\frac{i}{\eps}\left(\Ee_n-\Ee_m\right)$ 
and
 $\mu_{nm}=\frac{1}{2}\left(\Ee_n+\Ee_m\right)$,  respectively.
\end{thm}   

\medskip

Therefore,  $\Phinm(x,k)$ satisfy the   eigenvalue problems  
\begin{equation}\label{eig1} 
\Le\Phi^{\eps}_{nm}(x,k)=\lambda_{nm}\Phi^{\eps}_{nm}(x,k)
\end{equation}

\begin{equation}\label{eig2}
\Me\Phi^{\eps}_{nm}(x,k)=\mu_{nm}\Phi^{\eps}_{nm}(x,k) \ ,
\end{equation}
in phase space $L^2(\R_{xk}^2)$

\medskip
\noindent
{\bf Remark 1.}
It is very important to note that for the computation of Moyal functions, directly in  phase space, we need both eigenvalue problems, (\ref{eig1}) and (\ref{eig2}), as it has been explained  in \cite{CFZ}, \cite{KP}. Moreover, there is  no evolution equation in phase space which corresponds  to the eigenvalue equation  (\ref{eig2}) and which could be deduced from  Schr\"{o}dinger formulation , as   is the case for the quantum Liouville equation. This means that (\ref{eig2}) cannot result naturally from some initial value problem governing the Wigner function. In order to derive  the second eigenvalue equation directly from phase space, Fairlie \& Manogue  \cite{FM} extended the Wigner function by introducing an imaginary time variable $s$, thus constructing a second initial value problem, with time derivative  $i\partial_s$  and space operator $\Me$, for the extended Wigner function. Both the mathematical role and the physical content of this new function are still to be understood.

\medskip
\noindent
{\bf Remark 2.}
 It is also important to note that it is not possible to compute the   limits of the  Moyal eigenfunctions  $\Phi^{\eps}_{nm}$  as $\eps\to 0$, for any  $n,m$ independent of each other, a situation  which can be  somehow considered as  a consequence of the Bohr-Sommerfeld quantisation rule.  This situation is a fundamental  obstruction for the computation of   the limit of  the  phase-space eigenfunction  expansion of the Wigner function the  $\eps\to 0$,  from which one would expect to obtain   an analogous generalised expansion of the solution of the classical Liouville  equation. What can be evaluated is the classical limit  $\eps\to 0$, 
 when $\ n,m \to \infty$ and $n-m=$ constant.
For integrable Hamiltonians, Berry \& Balazs  \cite{BE} \cite{BB} have computed the classical limit of Moyal functions $\Phinm$  in the case  $n=m$, which reads as $\Phi^0_{nn}(I,\theta)=\delta\left(H(I)-(E^0_n)\right)$, where $H(I)$ is the Hamltonian in action angle variables 
$(I,\theta)$. In the "simplest" case of the the harmonic oscillator, Ripamonti  \cite{RN}  and   Truman \& Zhao  \cite{TZ} have given independent  proofs for the classical limit of the corresponding Moyal eigenfunctions  $\Phinm$ for all $n,m=0,1,\dots$,  based on  the asymptotics of Laguerre polynomials. Finally,  a formal computation in \cite{WB} shows that the classical limit of Moyal eigenfunctions,  in terms of action-angle variables ,  and for all $n \neq m$, reads as
$$\Phi^0_{nm}(I,\theta)=e^{-i\frac{E^0_{nm}}{\gamma_{nm}}\theta}\delta\left(H(I)-\frac12(E^0_n+E^0_m)\right) \ .$$
where $ E^0_{n}=lim_{\eps \rightarrow 0} \frac{\Ee}{\eps}$, 
$ E^0_{nm}=E^0_{n}-E^0_{m}$,  and   $\gamma_{nm}=H'\left(H^{-1}(\frac12(E^0_n+E^0_m))\right)$.
Also from this formal computation  becomes evident the necessity of both eigenvalue equations (\ref{eig1}) and  (\ref{eig2}).

\bigskip
\noindent
\subsection{Harmonic approximation of Moyal eigenfunctions}
\bigskip

We proceed now to the construction an asymptotic expansion of the Moyal eiegenfunctions, for small $\eps$, starting from the so-called harmonic approximation of the eigenfunctions of   the  Schr\"odinger  operator in the configuration space. 

It is known that  the  eigenfunctions and eigenvalues  of the Schr\"odinger operator $\Hop$ can be approximated by the corresponding ones of an appropriate harmonic oscillator, provided that the  potential $V(x)$  satisfies the following conditions (\cite{HS}, \cite{BS})

\medskip
\noindent
(i) $V \in C^{\infty}(\R) $\\
(ii) $V\geq  0, \textrm{ for some }  R>0
\mathop {\inf }\limits_{|x| > R}V(x) >0 \ ,$\\
(iii) $V' \textrm{ has finite number of zeros } \{x^{(\alpha)}\}_{\alpha=1}^\kappa  \ ,$\\
(iv)$ \textrm{ for each }x^{(\alpha)}\ \     V''(x^{(\alpha)})>0 \ ,$\\ 
(v) $V \textrm{ polynomially bounded } |V(x)|\le c(1+|x|^m) \ .$

For simplicity we consider only the case   $\kappa=1$, we adopt the normalization  $x^{(1)}=0$ and, without loss of generality, we also assume that $V(0)=0$. Although asymptotics of eigenvalues and eigenfunctions are also available for multiple wells, in order to deal with this case it is necessary to consider detailed information on the decay of the eigenfunctions (see, e.g., \cite{CODUSE} and the references therein) and take account of tunnelling effects for the Wigner function which is a rather complicated task \cite{BAVO}.

Then, the eigenvalues $\Ee_n$ and the eigenfunctions $\ue_n(x),\ n=0,1,\dots \ ,$ have the asymptotic expansions 
\begin{equation}\label{ass1}
\frac{2}{\eps}\Ee_n=e_n+\sum\limits_{l=1}^{N}a_n^{(l)}\eps^l+O\left(\eps^{N+1}\right) \ ,
\end{equation}
and
\begin{equation}\label{ass2}
\eps^{\frac14}\ue_n(x)\sim \psi_{n}(\frac{x}{\sqrt{\eps}})+ \sum\limits_{l=1}^{\infty} \eps^{\frac{l}{2}}\psi_n^{(l)}(\frac{x}{\sqrt{\eps}}) \ ,  \ \  \psi_n^{(l)}\in L^2(\R), n=0,1,\dots,\ \ l=1,2,\dots \ ,
\end{equation}
respectively, where 
$e_n$, $\psi_n(x)$ are the eigenvalues and eigenfunctions of  the  harmonic oscillator 
 $\Hop_h:=-\triangle+x^2 $, 
 \begin{equation}
\begin{array}{l}
e_n=2n+1,\ \ \\
\psi_n(x)=(2^nn!\sqrt{\pi})^{-\frac12}e^{-\frac{x^2}{2}}H_n(x),\
n=0,1,2,..
\end{array}
\end{equation}
$H_n(x)$  being the  Hermite polynomials \cite{TH}. Hence we refer to the expansions (\ref{ass1}), (\ref{ass2}) as the harmonic approximation.

Note that (\ref{ass2}) is understood in the sense that
\begin{equation}
||[U_\eps^{-1}\ue_n-\sum\limits_{l=0}^{N}\eps^{\frac{l}{2}}\psi_n^{(l)}||_{L^2(\R)}=O(\eps^{(N+1)/2}) \ .
\end{equation}
where  $U_\eps$ is a unitary dilation of $f\in L^2(\R)$ defined by
 $$
\left[U_{\eps}f\right](x):=\eps^{-\frac14}f\left(\frac{x}{\sqrt{\eps}
}\right),\ x\in \R, \ .
$$

The coefficients $a_n^{(l)},\psi_n^{(l)}, n=0,1,\dots \ , \ l=1,2,\dots \ ,$ can be computed in closed form by the Rayleigh-Schr\"odinger perturbation technique \cite{RSIV}.

By substituting  the  asymptotic expansions  (\ref{ass2}) of the eigenfunctions $\ue_n$ into th definition
 of the Moyal eigenfunctions $\Phinm(x,k),\ n,m=0,1,\dots$, we get the formal expansions 

\begin{equation}\label{asf1}
\Phinm(x,k)\sim \Psinm(x,k) + \sum\limits_{l=1}^{\infty}\eps^{\frac{l}{2}}\Znm^{\eps,(l)}(x,k) \\
\end{equation}
The first term of the expansion  the
Moyal eigenfunction of the  harmonic oscillator $V_h(x)=x^2/2$, that is
$$
\Psinm(x,k):=\frac{1}{\pi\eps}\int_{\mathbf{R}}e^{-i\frac{2k}{\eps}\sigma}
\psie_{n}(x+\sigma)\overline{\psie_{m}(x-\sigma)}d\sigma,\
n,m=0,1,2,\dots
$$
with $\psi^{\eps}_n(x)=\left[U_{\eps}\psi_n\right](x)$,
and  $Z_{nm}^{\eps,(l)}$  given by
\begin{equation}\label{znm}
\begin{array}{l}
Z_{nm}^{\eps,(l)}(x,k)=[U_{\eps,(xk)}\tZnm^{(l)}](x,k)\\
\tZnm^{(l)}(\cx,\ck)=\sum\limits_{\mu=0}^{l}W[\psi_n^{(\mu)},\psi_m^{(l-\mu )}](\cx,\ck) \ .
\end{array}
\end{equation}
Here $W[f,g](x,k)=\frac{1}{\pi}\int_{\R}e^{-i2k\sigma}f(x+\sigma)\overline{g}(x-\sigma)d\sigma$ is the  cross-Wigner transform for $\eps=1$, and  $U_{\eps,(xk)}$ is a dilation  of $f\in L^2(\R_{xk}^2)$ in phase space, which is defined by
\begin{equation}\label{dilph}
\left[U_{\eps,(xk)}f\right](x,k):=\eps^{-1}f\left(\frac{x}{\sqrt{\eps}},\frac{k}{\sqrt{\eps}
}\right) \ ,  \ \ (x,k) \in \R_{xk}^2 \ ,
\end{equation}
and it has the property $\|\left[U_{\eps,(xk)}f\right]\|_{L^2}=\frac{1}{\sqrt{\eps}}\|f\|_{L^2}$.
Since the first term of the expansion (\ref{ass1}) pertains to the harmonic oscillator, we refer to  it as the {\it harmonic expansion (or harmonic approximation)} of the Moyal eigenfunctins.

For simplifying the formulae,  in the sequel we introduce the notation ${\widetilde f}(\cx,\ck)=[U^{-1}_{\eps,(xk)}f](\cx,\ck)$.

The asymptotic expansion (\ref{asf1}) of $\Phinm(x,k)$  is written in the scaled variables $(\cx,\ck)$  as
\begin{equation}\label{asf2}
\left[U^{-1}_{\eps,(xk)}\Phi^{\eps}_{nm}\right](\cx,\ck) \sim \Psi_{nm}(\cx,\ck) + \sum\limits_{l=1}^{\infty}\eps^{\frac{l}{2}}\tZnm^{(l)}(\cx,\ck)\\
\end{equation}
since  $ \Psinm(x,k)=[U_{\eps,(xk)}\Psi_{nm}](x,k)$, where  $\Psi_{nm}(x,k)$, are the  cross-Wigner transform of Hermite functions $\psi_{n}$,  
\[
\Psi_{nm}(x,k):=\frac{1}{\pi}\int_{\R}e^{-i2k\sigma}
\psi_{n}(x+\sigma)\overline{\psi_{m}(x-\sigma)}d\sigma. \]

Furthermore, by substituting the expansion (\ref{asf1}) into the eigenvalue equations (\ref{eig1}) and (\ref{eig2}), and equating the coefficients of same powers   of $\eps$ as it is customary in regular perturbation schemes,  we expect to obtain  a hierarchy of  equations for the coefficient functions $Z_{nm}^{\eps,(l)}$. This procedure is quite cumbersome since the operators $\Le$ and $\Me$ depend also on the parameter $\eps$. The first step is to rescale the eigenvalue problems  by  using the transform  $U_{\eps,(xk)}$ and then  we use the properties of the potential to expand appropriately  the phase space pseudo-differential operators.

Applying the transform $U_{\eps,(xk)}$  onto the  eigenvalue problems (\ref{eig1}), (\ref{eig2})\footnote{this amounts to the change of variables
$(x,k)\rightarrow (
\cx=\frac{x}{\sqrt{\eps}},\ \ck=\frac{k}{\sqrt{\eps}}$)}, we get ,
\begin{equation}\label{eig11}
{\widetilde L}^{\eps}U^{-1}_{\eps,(xk)}\Phi^{\eps}_{nm}=\frac{i}{\eps}(\Ee_n-\Ee_m
)U^{-1}_{\eps,(xk)}\Phi^{\eps}_{nm} 
\end{equation}
and
\begin{equation}\label{eig22}
{\widetilde M}^{\eps}U^{-1}_{\eps,(xk)}\Phi^{\eps}_{nm}(x,k)=\frac12(\Ee_n+\Ee_m )U^{-1}_{\eps,(xk)}\Phi^{\eps}_{nm}(x,k) \ .
\end{equation}
The operators ${\widetilde L}^{\eps}$ and ${\widetilde M}^{\eps}$ are derived from $\Le$ and $\Me$  are derived by conjugation with the phase-space dilation 

$${\widetilde L}^{\eps}:=[U_{\eps,(xk)}]^{-1} \Le [U_{\eps,(xk)}] \ ,$$

$${\widetilde M}^{\eps}:= [U_{\eps,(xk)}]^{-1} \Me [U_{\eps,(xk)}]. $$

Note that for smooth potential  $V(x)$, we can also use the expansions
\begin{equation*}
 \Le= k\frac{\partial}{\partial x}-V'(x)\frac{\partial}{\partial k}-
\sum_{j=1}^{\infty}\eps^{2j}\left(\frac{i}{2}\right)^{2j}\frac{V^{(2j+1)}(x)}{(2j+1)!}\frac{\partial^{(2j+1)}}{\partial k^{2j+1}}
\end{equation*}
and
\begin{equation*}
 \Me= -\frac{\eps^2}{8}\triangle_{xk}+H(x,k)
+\sum_{j=1}^{\infty}\eps^{2j}\left(\frac{i}{2}\right)^{2j}\frac{V^{(2j)}(x)}{(2j)!}
\partial^{(2j)}_k+\frac{\eps^2}{8}\partial^2_k \ ,
\end{equation*}
to get the corresponding expansions of 
 ${\widetilde L}^{\eps}$ and ${\widetilde M}^{\eps}$. These read as 
\begin{equation}\label{aplt}
 {\widetilde L}^{\eps}=
L_h+(\cx-\frac{1}{\sqrt{\eps}}V'(\sqrt{\eps}\cx))\pk-\sum\limits_{j=1}^{\infty}
\eps^j
\left(\frac{i}{2}\right)^{2j}\frac{1}{(2j+1)!}\frac{V^{(2j+1)}(\sqrt{\eps}\cx)}{\sqrt{\eps}}\pkj 
\end{equation}

\begin{equation}\label{apmt}
\frac{1}{\eps}{\widetilde M}^{\eps}=
M_h+(\frac{1}{\eps}V(\sqrt{\eps}\cx)-\frac{\cx^2}{2})+\sum\limits_{j=1}^{\infty}
\eps^j\left(\frac{-1}{4}\right)^j\frac{1}{(2j)!}V^{(2j)}(\sqrt{\eps}\cx)\pkkj
+\frac{1}{8}\pkk
\end{equation}
where
\begin{equation}
L_h=\ck \px-\cx\pk \ , \ \ M_h=-\frac18 \triangle_{\cx\ck} +\frac{\ck^2}{2}+\frac{\cx^2}{2} \ ,
\end{equation}
are the   dilations of the operators  
\begin{equation}\label{hlm}
\LH:=\Le=k\frac{\partial}{\partial x}-x\frac{\partial}{\partial k} \ , \  \ \ \MH:=\Me =-\frac18 \triangle_{ x k} +\frac{ k^2}{2}+\frac{ x^2}{2}
\end{equation}
pertaining to the harmonic oscillator $V_h(x)=x^2/2$.  

Using the smoothness assumptions of the potential we can further approximate ${\widetilde L}^{\eps}$ and ${\widetilde M}^{\eps}$ by

\begin{equation}\label{apl}
{\widetilde L}^{\eps}\sim L^{\eps} := L_h+\sum\limits_{\nu=1}^{\infty} \eps^{\frac{\nu}{2}}
\B_{\nu} ( \cx,\pk )
\end{equation}
and
\begin{equation}\label{apm}
 \frac{1}{\eps}{\widetilde M}^{\eps}\sim \frac{1}{\eps}M^{\eps}:= M_h+\sum\limits_{\nu=1}^{\infty}
\eps^{\frac{\nu}{2}}\Gamma_{\nu}( \cx,\pk)
\end{equation}
where
\begin{equation}
\begin{array}{l}
\B_{\nu}( \cx,\pk)=- V^{(\nu+2)}(0)\sum\limits_{j=0}^{[(\nu-1 )
/2]+1}
\left(\frac{i}{2}\right)^{2j}\frac{1}{(2j+1)!}\frac{\cx^{\nu+1-2j}}{(\nu+1-2j)!}\pkj,\
\ \nu\geq 1,\\
 \Gamma_{\nu} ( \cx,\pk )=V^{( \nu+2
)}(0)\sum\limits_{j=0}^{[\nu/2]+1}\left(\frac{i}{2}\right)^{2j}
\frac{1}{(2j)!}\frac{\cx^{\nu+2-2j}}{(\nu+2-2j)!}\pkkj,\ \ \nu\geq
1 \ ,
\end{array}
\end{equation}
with $V^{(j)}(0)\equiv \frac{\partial^{j} V}{\partial x^{j}}(0)$.

Note that for polynomial potential $V(x)$, the above expansions are finite and exact, that is $L^{\eps}\equiv {\widetilde
L}^{\eps}$ and $M^{\eps}\equiv {\widetilde M}^{\eps}$.

Now, by substituting the expansions of  $ {\widetilde L}^{\eps}, \ {\widetilde M}^{\eps}$ , and also the expansions (\ref{ass1}) of the of eigenvalues 
$\Ee_{n}$ and of the eigenfunctions $U^{-1}_{\eps,(xk)}\Phi^{\eps}_{nm}$ (\ref{asf2}), into the scaled  eigenequations (\ref{eig11}), (\ref{eig22}), we obtain the following hierarchy of  non homogeneous problems for the correctors  $\tZnm^{(l)}\ n,m=0,1,\dots, \ l\geq
1$
\begin{eqnarray}\label{zeig}
\left[L_h-\frac{i}{2}(e_n-e_m)\right]\tZnm^{(l)}(\cx,\ck)=B_{nm}^{(l)}(\cx,\ck), \\
\left[M_h-\frac14(e_n+e_m)\right]\tZnm^{(l)}(\cx,\ck)=G_{nm}^{(l)}(\cx,\ck) \nonumber
\end{eqnarray}
where the right hand sides of (\ref{zeig}) are given by

\begin{equation}
\begin{array}{l}
B_{nm}^{(l)}(\cx,\ck)=-\sum\limits_{\nu=1}^{l}\B_{\nu}(\cx,\pk)\tZnm^{(l-\nu)}(\cx,\ck)
+\frac{i}{2}\sum\limits_{\nu=1}^{[l/2]}(a_n^{(\nu )}-a_m^{(\nu )})\tZnm^{(l-2\nu)}(\cx,\ck) \ , \\
G_{nm}^{(l)}(\cx,\ck)=-\sum\limits_{\nu=1}^{l}\Gamma_{\nu}(\cx,\pk)\tZnm^{(l-\nu)}(\cx,\ck)
+\frac14\sum\limits_{\nu=1}^{[l/2]}(a_n^{(\nu )}+a_m^{(\nu )})\tZnm^{(l-2\nu)}(\cx,\ck) \ ,  \\
\tZnm^{(0)}(\cx,\ck):= \Psi_{nm}(\cx,\ck) \ .
\end{array}
\end{equation}

It  can be shown by direct computation  that the functions $\tZnm^{(l)}(\cx,\ck)$ ( $n,m=0,1\dots$ and $l=0,1\dots$.)  given by (\ref{znm}) ,   satisfy the equations (\ref{zeig}).

The asymptotic expansion for the Moyal functions satisfies the $L^2$-estimate 
\begin{equation}\label{est1}
||\left[U_{\eps,(xk)}^{-1}\Phi^{\eps}_{nm}\right]-\sum
\limits_{l=0}^{N}\eps^{\frac{l}{2}}\tZnm^{(l)}||_{L^2(\R^2)}=O(\eps^{(N+1)/2})  \ .
\end{equation}
The proof is straightforward by using known estimates for the harmonic approximation of $ \ue_n$ (see Appendix A1).

\noindent
\section{The harmonic expansion of he  Wigner function}
\setcounter{equation}{0}
\renewcommand{\theequation}{3.\arabic{equation}}

In this section we use  the harmonic expansions for the Moyal eigenfunctions which were constructed in the previous section,  for the construction of an asymptotic expansion of the time-dependent Wigner function $\We(x,k,t)$ (recall the initial value problem (\ref{tql}). The first term of this expansion  is the Wigner function for the harmonic oscillator, and for this reason we call it {\it harmonic expansion} of the Wigner function. This asymptotic expansion suggests a harmonic  ansatz  which is then used for the construction of asymptotic expansions of the Wigner equation through a regular perturbation scheme directly in the phase space

First  we apply the Wigner transform onto the eigenfunction series solving the problem (\ref{Schro}) and we obtain an eigenfunction series of the Wigner function, in terms of the Moyal eigenfunctions. Then, we proceed formally and we approximate the coefficients and the Moyal eigenfunctions by their harmonic approximations.   It is important to  note that this expansion  is "quasi-asymptotic", since in general,  the coefficient  depend on the small parameter $\eps$ (and for this reason it is not a genuine semiclassical expansion).

\subsection{The eigenfunction expansion of the  Wigner function $\We(x,k,t)$}

Applying the  Fourier method,  we write the solution  $\psie(x,t)$ of the initial value problem ($\ref{Schro}$) for the Schr\"odinger equation as an eigenfunction series, in terms of the  eigenfunctions $\ue_{n}$ of the operator  $\Hop$ (see Section 2),  This reads as follows
\begin{equation}\label{exppsi}
\psie(x,t)=\sum_{n=0}^{\infty}\Anoe
\ue_{n}(x)e^{-i\frac{\Ee_n}{\eps}t} \ .
\end{equation}

The coefficients $\Anoe$ are given as the projection of initial data onto the eigenfunctions
\begin{equation}\label{exppsicoef}
\Anoe=(\psie_0,\ue_{n})_{L^2(\R)}
\end{equation}

By taking the Wigner transform
$$
\We (x,k,t) = \frac{1}{2\pi} \int_{-\infty}^{\infty} e^{-ik\sigma}
\psie \left(x+\frac{\eps\sigma}{2},t\right)\cpsie \left(x-\frac{\eps\sigma}{2}
,t\right)d\sigma \ ,
$$
of (\ref{exppsi}),
and using the definition (\ref {phinm}) of the Moyal eigenfunctions $\Phi^{\eps}_{nm}$, we obtain the following eigenfuction expansion of the Wigner function  
\begin{equation}\label{expw}
\We(x,k,t)=\sum_{n=0}^{\infty}\sum_{m=0}^{\infty}A_{nm}^{\eps}(t) \Phi^{\eps}_{nm}(x,k) \ , \\
\end{equation}
where
\begin{equation}\label{expwcoef1}
A_{nm}^{\eps}(t)=\frac{(\We_0,\Phinm )_{L^2(\R^2_{xk})}}{\parallel\Phinm\parallel_{L^2(\R^2)}^2}e^{-i\frac{\Ee_n-\Ee_m}{\eps}t}=
\eps (\We_0,\Phinm )_{L^2(\R^2_{xk})}e^{-i\frac{\Ee_n-\Ee_m}{\eps}t}
\end{equation}

It is easy to see that the coefficients   (\ref{expwcoef1}) are related to coefficients (\ref{exppsicoef}) by the relation
\begin{equation}\label{expwcoef2}
A_{nm}^{\eps}(t)=\Anoe\overline{\Amoe}e^{-i\frac{\Ee_n-\Ee_m}{\eps}t} \ .
\end{equation}

 The  coefficients  $A_{nm}^{\eps}(t),\ \ n,m=0,1\dots$  are approximated by combining (\ref{exppsicoef}) with the asymptotic expansions (\ref{ass1})  of the eigenvalues $\Ee_n$ and (\ref{ass2}) of the 
  Schr\"odinger eigenfunctions $\ue_{n}$, and they have the expansion
 \begin{equation}\label{anm}
A_{nm}^{\eps}(t)\sim
A_{h,nm}^{\eps}(t)+\sum\limits_{j=1}^{\infty}\eps^{\frac{j}{2}}\Delta_{j,nm}^{\eps}(t)
\end{equation}
where
\begin{equation}
A_{h,nm}^{\eps}(t)=
\eps(\We_0,\Psinm )_{L^2(\R^2_{xk})}e^{-i\frac{e_n-e_m}{2}t} \ ,
\end{equation}
and
\begin{equation}
\begin{array}{l}
\Delta_{1,nm}^{\eps}(t)=e^{-i\frac{e_n-e_m}{2}t}\left( [U_{\eps,(xk)}^{-1} \We_0 ],\tZnm^{(1)}\right)_{L^2(\R^2_{\cx \ck})} \ ,\\
\Delta_{2,nm}^{\eps}(t)=e^{-i\frac{e_n-e_m}{2}t}\left[\left( [U_{\eps,(xk)}^{-1} \We_0 ],\tZnm^{(2)}\right)_{L^2(\R^2_{\cx \ck})}+(-it)(a_n^{(1)}-a_m^{(1)}) \left( [U_{\eps,(xk)}^{-1} \We_0 ],\Psi_{nm}\right)_{L^2(\R^2_{\cx \ck})}\right] \ ,\\
\Delta_{3,nm}^{\eps}(t)=e^{-i\frac{e_n-e_m}{2}t}\left[\left( [U_{\eps,(xk)}^{-1} \We_0 ],\tZnm^{(3)}\right)_{L^2(\R^2_{\cx \ck})}+(-it)(a_n^{(1)}-a_m^{(1)}) \left( [U_{\eps,(xk)}^{-1} \We_0 ],\tZnm^{(1)}\right)_{L^2(\R^2_{\cx \ck})}\right] \  \dots  \ ,
\end{array}
 \end{equation}

Furthermore, by substituting the approximations (\ref{anm}) and  (\ref{asf1}) of the coefficients  $A_{nm}^{\eps}(t)$ and the eigenfunctions $\Phinm$, respectively, into the eigenfunction expansion  (\ref{expw}), we obtain the following expansion of the Wigner function 
\begin{equation}\label{qlas1}
\We(x,k,t)\sim \We_h(x,k,t)+\sum_{l=1}^{\infty} \eps^{l/2} Z^{\eps,(l)}(x,k,t) \ ,
\end{equation}
IBy direct computation we see that the function 
\begin{equation}\label{asexpw}
\We_h(x,k,t)=\sum_{n=0}^{\infty}\sum_{m=0}^{\infty}A_{h,nm}^{\eps}(t)\Psinm(x,k) \ .
\end{equation}
 satisfies  the initial value problem
\begin{equation}\label{hl}
\left\{
\begin{array}{l}
\frac{\partial}{\partial t}\We_h(x,k,t)+\LH\We_h(x,k,t)=0, (x,k)\in\R^2, t>0 \\\
\We(x,k,t)|_{t=0}=\We_0(x,k) \ ,
\end{array}
\right.
\end{equation}
with  (recall (\ref{hlm}))
$$\LH=k\frac{\partial}{\partial x}-x\frac{\partial}{\partial k} \ ,$$
 which  governs the evolution of the Wigner function for the  harmonic oscillator, with the same initial data $\We_0(x,k)$ (recall the problem (\ref{tql}) governing $\We$).

\bigskip
\subsection{The harmonic ansatz}

We pretend now that we do not know anything about the Schr\"odinger formulation, and we want to use the harmonic expansion (\ref{asexpw}) as an approximate solution (harmonic ansatz) to construct an approximate solution of the Wigner equation  (\ref{tql}).

In order to   construct the  equations for  the coefficient  $Z^{\eps,(l)}$,  we   apply the dilation  $ U^{-1}_{\eps,(xk)}$  defined by (\ref{dilph}), both on the problem (\ref{tql}) and  on the expansion  (\ref{qlas1}). In the new variables $(\cx,\ck) $ the Wigner equation  becomes
 \begin{equation}\label{ww}
\left\{
\begin{array}{l}
\frac{\partial}{\partial t}\tWe(\cx,\ck,t)+{\widetilde
L}^{\eps}\tWe(\cx,\ck,t)=0\\
\tWe(x,k,t)|_{t=0}=\tWe_0(x,k) \ ,
\end{array}
\right.
\end{equation}
while the expansion reads
\begin{equation}\label{qlas2}
\tWe(\cx,\ck,t)\sim \tWe_h(\cx,\ck,t)+ \sum_{l=1}^{\infty}\eps^{l/2}{\widetilde Z}^{\eps,(l)}(\cx,\ck,t) \ .
\end{equation}

Substituting the transformed expansion (\ref{qlas2}) of  $\tWe(\cx,\ck,t)$ and the expression (\ref{aplt}) of the operator ${\widetilde L}^{\eps}$  into (\ref{ww}), and then equating coefficients of same powers of $\eps$, we obtain the following  initial value problems 
\begin{equation}\label{wh}
\left\{
\begin{array}{l}
\frac{\partial}{\partial t}\tWe_h(\cx,\ck,t)+L_h\tWe_h(\cx,\ck,t)=0 \\
\tWe_h(\cx,\ck,t)|_{t=0}=\tWe_0(\cx,\ck)\ ,
\end{array}
\right.
\end{equation}
for the "harmonic term" $\tWe_h(\cx,\ck,t)$, and the following hierarchy of problems 
\begin{equation}\label{zl}
\left\{
\begin{array}{l}
\frac{\partial}{\partial t}{\widetilde Z}^{\eps,(l)}(\cx,\ck,t)+L_h{\widetilde Z}^{\eps,(l)}(\cx,\ck,t)=D^{\eps,(l)}(\cx,\ck,t),\ l\geq 1\\
{\widetilde Z}^{\eps,(l)}(\cx,\ck,t)|_{t=0}=0 \ ,
\end{array}
\right.
\end{equation}
for the higher-order coefficients${\widetilde Z}^{\eps,(l)}(\cx,\ck,t),\ l\geq 1$. The right hand side  
$D^{\eps,(l)}(\cx,\ck,t$ of (\ref{zl}) is given by 
\[
D^{\eps,(l)}(\cx,\ck,t)=-\B_{l}(\cx,\pk)\tWe_h(\cx,\ck,t)-
\sum_{\nu=1}^{l-1}\B_{\nu}(\cx,\pk){\widetilde Z}^{\eps,(l-\nu)}(\cx,\ck,t)
\]

\medskip
\noindent
{\bf Remark 3 .} The initial data of the f problem for $\We_h$ are the same with those of the original problem (\ref{tql}),
and therefore  the higher order problems have zero initial data but they are forced from lower orders. By this choice we avoid to expand  the initial  functions $\tWe_0$ with respect to the small parameter $\eps$, which for a WKB initial wave function  $\psie_0$, leads to a distributional  expansion (compare with the distributional expansion constructed by Pulvirenti \cite{PU}). A  consequence of our choice is that  the coefficients   ${\widetilde Z}^{\eps,(l)}$ are $\eps$-dependent, and therefore the expansion (\ref{qlas1})  is not  a genuine   semi-classical expansion (see also the comments in \cite{FN1} for such type of expansions). Other choices of the initial data for the harmonic problem would be either the Airy approximation of $\tWe_0$  which has been proposed by Berry \cite{BE} (see also \cite{FM1}), or a  wavepacket expansion in phase space, that can be derived by applying the Wigner transform on the FBI transform of the initial wavefunction..These two alternative choices are still open to investigation.

The  problems (\ref{wh}) and(\ref {zl}) can be integrated by the method of characteristics. 
We demote bt  $g_h^t(q,p)$  the Hamiltonian flow of harmonic oscillator. This flow  is the solution of the Hamiltonian system
$$
\frac{d\xi}{dt}=\eta \ , \ \ \ \ \frac{d\eta}{dt}=-\xi \ ,
$$
with initial conditions
 $$\xi(t=0;q,p)=q,\ \  \eta(t=0;q,p)=p \ ,$$
and it is given by
\begin{align*}
g_h^t(q,p)&=\left(g_{h,1}^t(q,p),g_{h,2}^t(q,p)\right)\\
&=\left(\cx(q,p,t):=q\cos(t)+p\sin(t),\ck(q,p,t):=p\cos(t)-q\sin(t)\right) \ .
\end{align*}
The inverse flow is 
\begin{align*}
g_h^{-t}(\cx,\ck)&=\left(g_{h,1}^{-t}(\cx,\ck),g_{h,2}^{-t}(\cx,\ck)\right)\\
&=\Bigl(q(\cx,\ck,t):=\cx \cos(t)-\ck \sin(t), p(\cx,\ck,t):=\ck \cos(t)+\cx \sin(t)\Bigr)
\end{align*}
Then,  the solutions of the problems (\ref{wh}) and  (\ref{zl}) are given by the  formulae
\begin{equation}\label{solwh}
\tWe_h(\cx,\ck,t)=\We_0(q(\cx,\ck,t),p(\cx,\ck,t))
\end{equation}
and
\begin{equation}\label{solzh}
{\widetilde Z}^{\eps,(l)}(\cx,\ck,t)=\int_{0}^{t}D^{\eps,(l)}(q(\cx,\ck,t-s),p(\cx,\ck,t-s)),s)ds
\end{equation}
It is important to note that the dependence of  $\tWe_h(\cx,\ck,t)$ and
${\widetilde Z}^{\eps,(l)}(\cx,\ck,t)$ on $\eps$, comes  only from the dependence of initial function $\We_0$ on $\eps$.

The validity of expansion (\ref{qlas1}) depends crussialy on the properties of the initial data  $\We_0$. Recall here that the required properties of the potential have been stated at the beginning of Section 2.1. 

In order to understand the speciality of the WKB initial data, we will now consider two classes of initial data in phase space. The first class consists of those data whose $U^{-1}_{\eps,xk}$  dilation (recall eq (\ref{dilph})) is an $\eps$-independent function (in the scaled variables), and in this case  we can prove the validity of the expansion in   $L^2$ norm. The second class consists of initial data  which are the Wigner transform of a certain  WKB initial wave function, and in this case  we can  prove an estimate for the remainder in a weighted $L^2$ norm.

The proofs of the theorems stated below are straightforward applications  of the  technique proposed by Bouzouina \& Robert \cite{BR}. who intro cued it for proving  a novel (Egorov-type) estimate for the remainder in the  $L^2$ operator norm for the semiclassical expansion of the evolution of quantum observables. The proofs  are given in the Appendix A2.

\begin{thm}\label{thm1}
Let  $\tWe$ the solution of the initial value problem (\ref{ww}) with $\eps$-independent initial data $\widetilde{f}_0(\cx,\ck)\in\Sh(\R^2)$ .
Then, for any    $N\in \N$,
\begin{equation}\label{expthm2}
\tWe(\cx,\ck,t)= \tWe_h(\cx,\ck,t)+\sum_{l=1}^N \eps^{l/2} {\widetilde Z}^{\eps, (l)}(\cx,\ck,t)+R^{N+1}(\cx,\ck,t)
\end{equation}
where $ {\widetilde Z}^{\eps, (l)}$ are given by formulas (\ref{solzh}).
 Moreover the following bound for the remainder holds
 \begin{equation}\label{remthm2}
\parallel R^{N+1}(t)\|_{L^2(\R^2_{\cx\ck})} \le C_{N}e^{t} \eps^{(N+1)/2} \ 
\end{equation}.
\end{thm} $ \hfill \Box$

An interesting application of  Theorem 2 is in the case where the initial data $\widetilde{f}_0(\cx,\ck)$ are the dilated Wigner transform of a coherent state 
$$\psie_0(x)=\left(\frac{1}{\pi\eps}\right)^{1/4}e^{\frac{i k_0 x}{\eps}} e^{-\frac{(x-x_0)^2}{2\eps}} \ , $$ 
which is given by
$$ \widetilde{f}_0(\cx,\ck)= [U_{\eps,(xk)}^{-1} \We_0](\cx,\ck)=\frac{1}{\pi}e^{-(\cx-\cx_0)^2}e^{-(\ck-\ck_0)^2} \ ,$$
where $\cx_0=\frac{x_0}{\sqrt{\eps}},\ \ck_0=\frac{k_0}{\sqrt{\eps}}$. Then,  
$\widetilde{f}_0(\cx,\ck)\in \Sh(\R^2)$and it is independent of  $\eps$.

For a WKB initial wavefunction 
$$\psie_0(x)=\alpha_0(x)e^{i\frac{S_0(x)}{\eps}} \in L^2(\R) \ , $$
the dilated  Wigner function $\widetilde{f}_0(\cx,\ck)=\widetilde{f}^{\eps}_0(\cx,\ck)$ is $\eps-$dependent, and an $L^2$ estimate of the error leads to negative powers of $\eps$, a fact which is expected since the weak limit, as 
$\eps \rightarrow 0$, of the Wigner transform of a WKB state,  is a Dirac distribution. Thus the estimate of the remainder we provide here is based on a weighted $L^2$ norm with a Gaussian  weight
$$r^{\eps}(\cx,\ck)=e^{-\frac{\cx^2+\ck^2}{\eps^2}} \ .$$
In this case the theorem reads as follows.

\medskip
\begin{thm}\label{thm2}
Let $\tWe$ the solution of the initial value problem   (\ref{ww}) with initial data  $\widetilde{f}^{\eps}_0(\cx,\ck)= [U^{-1}\We_0](\cx,\ck)$, where  $\We_0$ is the Wigner transform of a WKB initial wavefunction  $\psie_0(x)=\alpha_0(x)e^{i\frac{S_0(x)}{\eps}}$, with
$$\alpha_0(x)=e^{-x^2/2} \ , \ \   S_0(x)=x^2/2 \ \ \ \text{or}\ \ S_0(x)=x \ .$$
Then for all  $N \in \N$,
\[\tWe(\cx,\ck,t)=\tWe_h(\cx,\ck,t)+ \sum_{l=1}^N \eps^{l/2} {\widetilde Z}^{\eps,(l)}(\cx,\ck,t)+R^{\eps,N+1}(\cx,\ck,t)\]
where  $ {\widetilde Z}^{\eps,(l)}$  are given by formulas (\ref{solzh}). Moreover the following bound for the remainder holds
$$ \|R^{\eps,N+1}(t)\|_ {L^2_{r^{\eps}}}\le C_{N}e^{t} \eps^{(N+1)/2} \ ,$$
where $\|\cdot \|_ {L^2_{r^{\eps}}}$ denotes the  $L^2_{r^{\eps}} $ norm,
$$\| f \|_ {L^2_{r^{\eps}}} =\left(\int_{\R^2} |f(\cx,\ck)|^2 r^{\eps}(\cx,\ck) d\cx d\ck\right)^{1/2} \ , $$
with $r^{\eps}(\cx,\ck)= e^{-\frac{\cx^2+\ck^2 }{\eps^2}}$.
\end{thm} $ \hfill \Box$

\bigskip
\section{The classical expansion of the Wigner function} 
\setcounter{equation}{0}
\renewcommand{\theequation}{4.\arabic{equation}}

When the potential is smooth,the quantum Liouville operator $\Le$,  can be expanded in the form  (see \cite{HWL}) 
 \begin{equation}
 \Le=\Lc-\sum_{j=1}^{\infty}\eps^{2j}\Theta_j (x,\frac{\partial}{\partial k}) \ ,
\end{equation}
where
\begin{equation}
\Lc\equiv k\frac{\partial}{\partial x}-V'(x)\frac{\partial}{\partial k} \ ,
\end{equation}
is the classical Liouville operator, and 
\begin{equation}
\Theta_j (x,\frac{\partial}{\partial k})\equiv \left(\frac{i}{2}\right)^{2j}\frac{V^{(2j+1)}(x)}{(2j+1)!}\frac{\partial^{(2j+1)}}{\partial k^{2j+1}} \ .
\end{equation}

We observe  that formally, as $\eps\rightarrow  0$, the operator   $\Le$ reduces to the classical operator $\Lc$. Therefore, it is plausible to assume the expansion \cite{ST}
\begin{equation}\label{qlas3}
\We(x,k,t)\sim \We_c(x,k,t)+ \sum_{l=1}^{\infty}\eps^{2l}Z_c^{\eps,(l)}(x,k,t) \ .
\end{equation}
where $\We_c(x,k,t)$ satisfies the classical problem
\begin{equation}\label{wc}
\left\{
\begin{array}{l}
\frac{\partial}{\partial t}\We_c(x,k,t)+\Lc\We_c(x,k,t)=0 \\
\We_c(x,k,t)|_{t=0}=\We_0(x,k) \ .
\end{array}
\right.
\end{equation}
For this reason, in the sequel we will refer to the expansion(\ref{qlas3}) as the {\it classical approximation}. This expansion has been proposed by Steinr\"uck \cite{ST} for the case of  $\eps$-independent initial data, and it has been later studied by F. Narcowich \cite{FN1}, and recently rigorously by  Pulvirenti \cite{PU} for a class of WKB initial data (see also the related studies\cite{AR}, \cite{ROB}).

Substituting  (\ref{qlas3}) into the problem (\ref{tql}),
and separating   powers of $\eps$ as it is customary in regular perturbations, we find that while $Z_c^{\eps,(l)}$ satisfy the hierarchy of problems
\begin{equation}\label{zcl}
\left\{
\begin{array}{l}
\frac{\partial}{\partial t}Z_c^{\eps,(l)}(x,k,t)+\Lc Z_c^{\eps,(l)}(x,k,t)=\Theta^{(l)}(x,k,t),\ l\geq 1\\
Z_c^{\eps,(l)}(x,k,t)|_{t=0}=0 \ ,
\end{array}
\right.
\end{equation}
where
$$\Theta^{(l)}(x,k,t)=\sum_{j=1}^{l}\Theta_{j} (x,\frac{\partial}{\partial k})Z_c^{\eps,(l-j)}(x,k,t) \ ,$$
and $Z_c^{\eps,(0)}\equiv \We_c$.

The solutions of problems (\ref{wc}) and (\ref{zcl}) are constructed by the method of characteristics, and they are given by 

\begin{equation}\label{clsol}
\begin{array}{l}
\We_c(x,k,t)=\We_0( q(x,k,-t), p(x,k,-t))\\
\\
Z_c^{\eps,(l)}(x,k,t)=\int_0^t \Theta^{(l)}(q(x,k,t-s) ,p(x,k,t-s) ,s)ds
\end{array}
\end{equation}
where $q(x,k,t)\ ,\ p(x,k,t))$ are the bicharacteristics associated to the potential $V$, that is the solutions of the Hamiltonian system
\begin{equation}\label{hams}
\left\{
\begin{array}{l}
\frac{dq}{dt}=p,\qquad \frac{dp}{dt}=-V'(q)\\
\\
q(0)=x,\qquad p(0)=k  \ .
\end{array}
\right.
\end{equation}

We must emphasise here that in the classical expansion (\ref{qlas3}). the initial data are propagated along the bicharacteristics associated to the  potential $V$, in contrary to the harmonic approximation developed in Section 3, where the initial data are propagated along the bicharacteristics of the approximating harmonic oscillator with potential $V_h$.

\newpage
\noindent
{\bf Remark 4. }(on the structure of the classical expansion)

\noindent
(a)The expansion (\ref{qlas3}) is not a genuine  semiclassical expansion because its coefficients depend on the small parameter $\eps$.

\noindent
(b) The  classical expansion is of the multiplicative type 
$$\We(x,k,t)= \We_c(x,k,t)\Bigl(1 +\sum_{\ell \ge 1} \eps^{\ell}\We_{\ell}(x,k,t)\Bigr) \ ,$$
which is not the case for the harmonic expansion.

\noindent
(c) As  $\ \eps \rightarrow 0$,  all terms of the classical expansion concentrate near the  Langrangian manifold  $\Lambda_{t}=\{\ p(x,k,t) = S'_{0}(q(x,k,t)\ \}$ generated by  the Hamiltonian flow. Therefore, in the case that the ray field has caustics, the first term $\ \We_c(x,k,t)\ $ of the expansion behaves much as a Dirac distribution,and it is not efficient in computing energy densities for fixed position in configuration space (cf \cite{FM1}).

\medskip
\noindent
{\bf Remark 5. }(on the  applicability of the expansion)

\noindent
It follows from the  construction of the approximations that the harmonic expansion is expected to be applicable  at least in a region  of width $O(\sqrt{\eps})$ near the bottom of the potential well, and the classical expansion  near the Lagrangian manifold,  both for  short time. For this reason, in the example of the next section  we compare the two expansions near the potential well of an anharmonic (quartic) oscillator.

\noindent
\section{ Example: Anharmonic oscillator and caustics}
\setcounter{equation}{0}
\renewcommand{\theequation}{5.\arabic{equation}}

As an application of the developed expansions, we combine the  harmonic and the classical  expansions  of the  Wigner  function with property  (\ref{ampl}), in order to estimate the amplitude $|\psie|$ of the wavefunction for the quartic oscillator with (anharmonic) potential
\begin{equation}
V(x)=x^2/2+\mu x^4/4 \ ,  \qquad \mu >0 \ ,
\end{equation}
and   WKB  initial data  (\ref{wkb0}) of the Gauss-Fresnel type, that is 
\begin{equation}
a_0(x)=e^{-\frac{x^2}{2}},\ S_0(x)=\frac{x^2}{2} \ .
\end{equation}

The corresponding Wigner equation in phase space is

\begin{equation}\label{ql}
\left\{
\begin{array}{l}
\frac{\partial}{\partial t}\We(x,k,t)+\Le\We(x,k,t)=0, (x,k)\in\R^2, t>0\\

\We(x,k,t)|_{t=0}=\We_0(x,k)
\end{array}
\right.
\end{equation}
where
\begin{equation}
\Le\equiv k\frac{\partial}{\partial x}-(x+\mu
x^3)\frac{\partial}{\partial k}+ \frac{\eps^2}{4}\mu x
\frac{\partial^{(3)}}{\partial k^3} \ ,
\end{equation}
with initial data
\begin{equation}
\We_0(x,k)=\frac{1}{\sqrt{\pi}
\eps}e^{-x^2}e^{-\frac{(k-x)^2}{\eps^2}} \ .
\end{equation}

First we compute the bicharacteristics and the rays for  the corresponding harmonic oscillator (with potential $V_h(x)=x^2/2$), and also the bicharacteristics of the quartic oscillator.

\medskip
\noindent
\paragraph{Bicharacteristics, rays and caustics.}

The bicharacteristics of the harmonic oscillator are easily computed from the Hamiltonian system
 \begin{equation}
\left\{
\begin{array}{l}
\frac{dx}{dt}=k,\ \ x(0)=q \\
\frac{dk}{dt}=-V'_{h}(x)=-x\ \ k(0)=p
\end{array}
\right.
\end{equation}
and they are given by
\begin{equation}
(x_h (q,p,t),k_h (q,p,t))=g_h^t(q,p)=(q\cos(t)+p\sin(t),p\cos(t)-q\sin(t)) \ .
\end{equation}
The inverse bicharactericts are
\begin{equation}
(q_h(x,k,t),p_h(x,k,t))=g_h^{-t}(x,k)=(x \cos(t)-k \sin(t),k \cos(t)+x \sin(t)) \ .
\end{equation}

From the condition  $p =S_0'(q)=q$, we obtain the equations of the rays  
\begin{equation}
\tilde{x}_h=\tilde{x}_h(t;q)=q(\cos(t)+\sin(t))
\end{equation}
and by solving the equation   
\begin{equation}
J(q,t)=\frac{\partial \tilde{x}_h}{\partial q}=(\cos(t)+\sin(t))=0 
\end{equation}
 with respect to  $q=q(t)$, we find the caustics, which for harmonic oscillator is a sequence of focal points (Figure 1)
\begin{equation}
(x_\nu, t_\nu)=(0, \nu\pi-\frac{\pi}{4}),\ \ \nu=1,2,...
\end{equation}

 \begin{figure}[ht1]
\centering
\includegraphics
[width=0.5\textwidth]{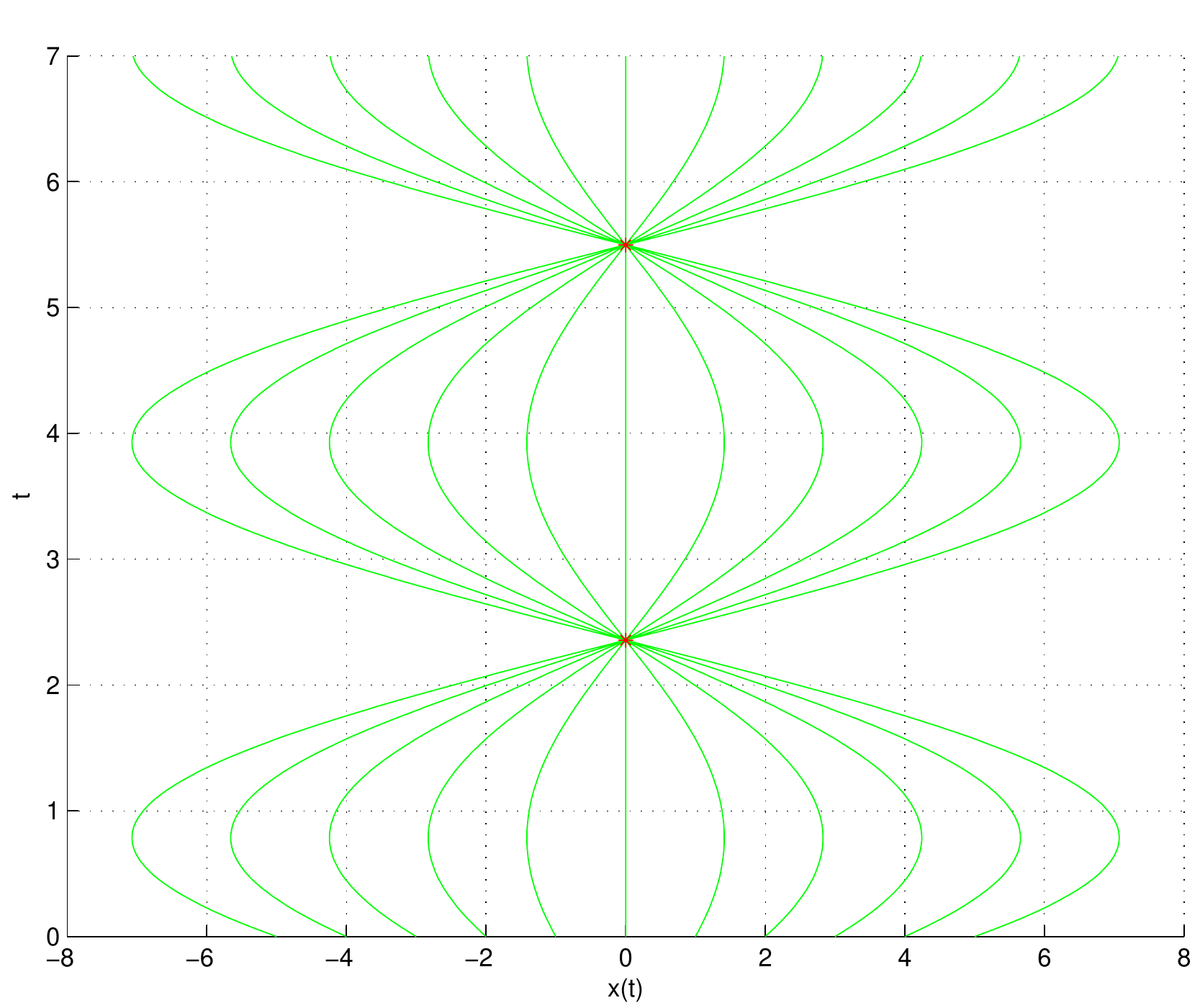}
\caption{{\it Rays \& caustic Harmonic oscillator $V_h(x)=x^2/2$}}
  \end{figure}

The bicharacteristics $(x_{V}(q,p,t), k_{V}(q,p,t))=g^{t}_{V}(q,p)$ for the quartic oscillator are found  from the corresponded Hamiltonian system, and they are given by
\begin{equation}
x_V(q,p,t)=A(q,p)sd(\Gamma(q,p)t+C(q,p),B(q,p)), \qquad
k_V(q,p,t)=\frac{\partial x_V}{\partial t}(q,p,t)
\end{equation}
where
\begin{equation}
\begin{array}{l}
A(q,p)= \frac{\sqrt{c(q,p)}}{( 2\mu c(q,p)+1 )^{1/4}} \ , \qquad \qquad
B^2(q,p)=\frac{\sqrt{2\mu c(q,p)+1}-1}{2\sqrt{2\mu c(q,p)+1}} \ ,\\
\\
\Gamma(q,p)=(2\mu c(q,p)+1)^{1/4} \ , \qquad \qquad
C(q,p)=sd^{-1}(\frac{q}{A(q,p)},B(q,p))
\end{array}
\end{equation}
with
$$c(q,p)=p^2+q^2+\mu\frac{q^4}{2} \ ,$$
and  $sd(a,b)=\frac{sn(a,b)}{dn(a,b)}$,$sn,\ dn$  are the  Jacobi elliptic functions.

Unfortunately it is not possible to obtain an analytical formula for the caustic. However, by considering the rays  $\tilde{x}_V=\tilde{x}_V(t;q)=x_V(q,p=S'_{0}(q),t)$ and solving numerically  the equation  $J(q,t)=\partial_{q}\tilde{x}_V=0$,  which is available in explicit form, we have observed that the caustic consists of  a family of cusps with beaks at the focal points of the corresponding harmonic oscillator (Figure 2).
We have checked analytically this observation by proving that the focal points of harmonic oscillator are indeed zeros of the Jacobian for the quartic oscillator.

 \begin{figure}[ht1]
\centering
\includegraphics[width=0.5\textwidth]{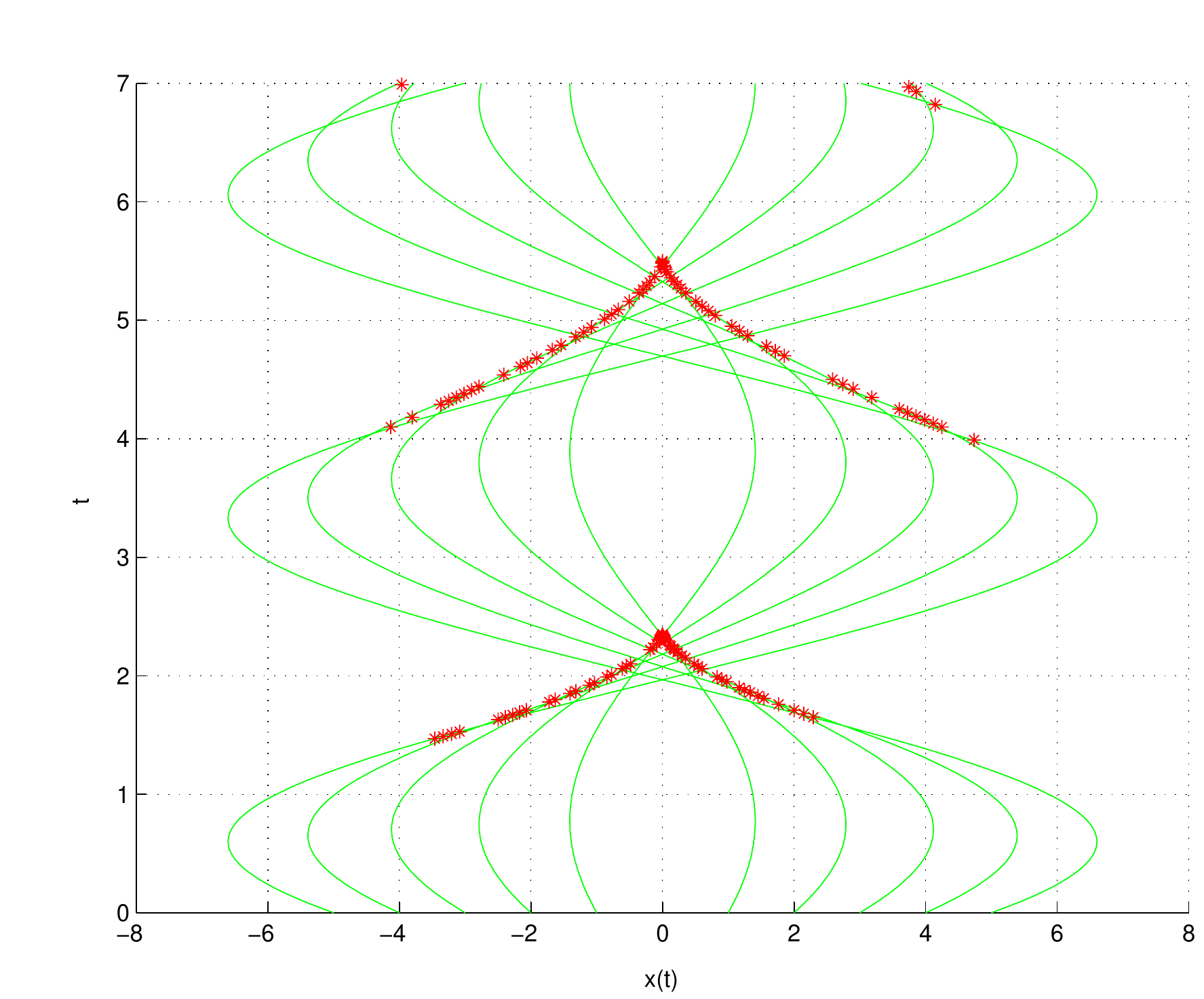}
\caption{{\it Rays \& caustic Quartic oscillator  $V(x)=x^2/2 +\mu x^4/4$}}
 \end{figure}

Since the expressions of the bicharacteristics for the anharmonic oscillator are very complicated, the analytical computation of coefficient $Z_c^{\eps,(l)}(x,k,t)$  of the classical expansion isimpossible. Nevertheless, it is possible to compute the classical expansion of the Wigner function approximately, by using an  approximation of the characteristics  for small values of the coupling constant $\mu$. Indeed, by  the method of multiple scales, for $\mu$ being the small parameter, we solve the ordinary differential equation 
$$\ddot{x}+ x +\mu x^3=0 \ ,  \ \ x(0)=q \ , \  \ \dot x(0)=p \ ,$$
which,   is equivalent to the Hamiltonian system  for the quartic oscillator, and  we get the following approximation of the bicharacteristics
\begin{equation}\label{quartbich}
\left\{
\begin{array}{l}
x_a(q,p,t)=q\cos(\omega t)+p\sin(\omega t)+O(\mu)\\
k_a(q,p,t)=\dot x_a(q,p,t))=p\omega\cos(\omega t)-q\omega\sin(\omega t)+O(\mu) \ ,
\end{array}
\right.
\end{equation}
and also the approximation of the inverse bicharacteristics
\begin{equation}
\left\{
\begin{array}{l}
q_a(x,k,t)=x_a(x,k,-t)=x\cos(\omega t)-k\sin(\omega t)+O(\mu)\\
p_a(x,k,t)=x_a(x,k,-t)=k\omega\cos(\omega t)+x\omega\sin(\omega t)+O(\mu) \ ,
\end{array}
\right.
\end{equation}
where $\omega= \omega(q,p;\mu)=1+\frac38\mu(q^2+p^2)+O(\mu^2)$ is the approximate angular velocity of the quartic oscillator.

\medskip
\noindent
\paragraph{Amplitude via harmonic approximation.}

The harmonic  expansion of $\We(x,k,t)$ reads  
\begin{equation}\nonumber
\We(x,k,t)\sim \We_h(x,k,t)+
\sum_{j=1}^{\infty}\eps^{j}Z^{\eps,(2j)}(x,k,t)
\end{equation}
where
\begin{eqnarray}
\We_h(x,k,t)&=&\We_0(q_h(x,k,t),p_h(x,k,t))=  \nonumber \\ 
&=&\frac{1}{\sqrt{\pi} \eps}e^{-(x\cos(t)-k\sin(t))^2}e^{-\frac{(k(\cos(t)+\sin(t))-x(\sin(t)-\cos(t)))^2}{\eps^2}} \ .
\end{eqnarray}
Integrating $\We_h$ with respect to $k$ (recall (\ref{ampl})), we compute the principal contribution of $\We(x,k,t)$  to  the amplitude
 \begin{eqnarray}
|\psie_h(x,t)|^2 \approx \int_{\R}\We_h(x,k,t)dk
=\frac{1}{\sqrt{\eps^2\sin^2(t)+(\cos(t)+\sin(t))^2}}\nonumber\\
\times exp\left(-x^2\frac{(-2\sin^2(t)+(\cos(t)+\sin(t))^2)^2}{\eps^2\sin^2(t)+(\cos(t)+\sin(t))^2}\right) \ .
\end{eqnarray}

At the focal points we have
\begin{equation}
|\psie_h(x=0,t_\nu)|^2 \approx \frac{\sqrt{2}}{\eps} \ .
\end {equation}

The coefficient $Z^{\eps,(2)}(x,k,t)$  are  computed in Appendix A3, in two different ways which lead to the same approximation up to the order $O(\eps^2)$. The  contribution of this term  in the amplitude $|\psie|^2$ at the focal points, is  also computed in the Appendx A3, and it is given by
\begin{equation}
\eps\int_{\R}Z^{\eps,(2)}(0, k,t_{\nu})d k=\eps\frac{1}{\sqrt{\eps}}\int_{\R}Z^{\eps,(2)}(0,\ck,t_{\nu})d\ck=\frac{\sqrt{2}}{\pi\eps}\mu (\beta+\beta^{\eps})
\end{equation}
where
$$\beta=\frac{\pi}{8} ( \mu-1/4)-3 \ , \qquad \beta^{\eps}=\frac{17}{2} \eps -(\frac{3\pi}{16}( \mu-1/4)+3)\eps^2 \ .$$
By continuing the computation to higher orders, it turns out  that  the contribution of the subsequent  terms of the expansion is  also of the order $O({1}/{\eps})$.

\medskip
\noindent
\paragraph{Amplitude via classical approximation.}
\medskip
The classical expansion (\ref{qlas3}) of $\We(x,k,t)$ reads 
\begin{equation*}
\begin{array}{l}
\We(x,k,t)\sim \We_c(x,k,t)+
\sum_{l=1}^{\infty}\eps^{2l}Z_c^{\eps,(l)}(x,k,t)\\
\end{array}
\end{equation*}
 The leading term $\We_c(x,k,t)$ and the subsequent coefficients  $Z_c^{\eps,(l)}(x,k,t)$ are  calculated using (\ref{clsol})   and  the approximate characteristics (\ref{quartbich}), and they are given by
 \begin{equation}
\begin{array}{l}
\We_c(x,k,t)\sim \We_a(q_a(x,k,t), p_a(x,k,t)):= \We_0(q_a(x,k,t), p_a(x,k,t)) \ , \\
\\
Z_c^{\eps,(l)}(x,k,t)= \int_0^t \Theta^{(l)}(q_a(x,k,s-t), p_a(x,k,s-t),s)ds \ .
\end{array}
\end{equation}

Integrating $\We_c$ with respect to $k$, we compute the principal contribution of $\We(x,k,t)$  to  the amplitude

\begin{align}
|\psie(0,t_\nu)|^2\approx&\int_{\R}\We_a(0,k,t_{\nu})dk\equiv \int_{\R}\We_0(q_a(0,k,t_{\nu}),p_a(0,k,t_{\nu}))dk \nonumber\\
= &\frac{1}{\sqrt{\pi}\eps}\int_{\R}e^{-\frac{k^2}{2}(1-\mu\gamma k^2)^2}e^{-(\frac{\mu}{\eps})^2\alpha^2k^6}dk\\
=&\frac{1}{\sqrt{\pi}\eps^{2/3}}\int_{\R}e^{-\frac{\eps^{2/3}\xi^2}{2}(1-\mu\gamma \eps^{2/3}\xi^2)^2}e^{-\mu^2\alpha^2\xi^6}d\xi \nonumber \\
=&\frac{1}{\sqrt{\pi}\eps^{2/3}\mu^{1/3}}\int_{\R}e^{-\frac{\eps^{2/3}y^2}{2\mu^{2/3}}(1-\mu^{1/3}\gamma
\eps^{2/3}y^2)^2}e^{-\alpha^2y^6}dy \ , \nonumber
\end{align}
where $\gamma,\ \alpha$ are constants independent of   $\mu$ and $\eps$.

Hence
\begin{equation}
|\psie(0,t_{\nu})|^2\sim O\left(\frac{1}{\eps^{2/3}\mu^{1/3}}\right),\  \textrm{for}\  \mu\sim \eps^{1-\delta},0\le \delta\le 1 \ ,
\end{equation}

and

\begin{equation}
|\psie(0,t_{\nu})|^2\sim O\left(\frac{1}{\eps}\right),\  \textrm{for}\  \mu\sim \eps^{1+\delta},   \ \ \delta \ge 0 \ .
\end{equation}

These estimates have  the same order with the corresponding one  which was computed from the harmonic expansion if we choose $\mu=O(\eps^{1+\delta})$. Therefore, if we allow the dependence of the coupling constant $\mu$ on the parameter $\eps$, the result shows that for this particular relation between the frequency and the strength of anharmonicity of the potential, both expansions predict the same semiclassical effect, at least at the focal points. However, from the above analysis we also expect agreement of the predicted wave fields everywhere.  Moreover, and more important, this picture suggests that different expansions should be used according to the actual relation of the parameters $\eps$ and $\mu$, but it is still open and quite difficult to determine  precise criteria for transferring from one approximation to the other.

Finally, it is interesting to remark that the constructed expansions, at least  for the example of quartic oscillator, are related as follows 
\begin{align}
\We(x,k,t)  \xrightarrow{{\eps<<1}} \We_c(x,k,t)+\sum \eps^{2l} Z_c^{\eps,(l)}\ \xrightarrow{{\mu<<1,O(\eps)}} \We_h(x,k,t)+ \sum \eps^l Z^{\eps,(2l)} \ .
\end{align}

\section{Discussion}
We have constructed a couple of  approximations of the Wigner function for the  Schr\"odinger equation with oscillatory initial data.  The first one, which we call the harmonic approximation, has the form
\begin{equation*}
\We(x,k,t)\sim \We_h(x,k,t)+
\sum_{j=1}^{\infty}\eps^{j}Z^{\eps,(2j)}(x,k,t)
\end{equation*}
where the  principal term $\We_h$ is the Wigner function of a harmonic oscillator associated to the harmonic approximation of the potential $V(x)$. The second one, which has been used  in quantum mechanics long time ago. has the multiplicative form
\begin{equation*}
\We(x,k,t)= \We_c(x,k,t)\Bigl(1 +\sum_{\ell \ge 1} \eps^{\ell}\We_{\ell}(x,k,t)\Bigr) \ ,
\end{equation*}
where $\We_c$ is the limit Wigner distirbution which is the solution of the Liouville equation of classical mechanics. For the construction of both approximations we choose  the initial data for the principal terms to be the initial Wigner function $\We(x,k,t=0)=\We_0(x,k)$, that is
$\We_h(x,k,t=0)=\We_0(x,k)$ and $\We_c(x,k,t=0)=\We_0(x,k)$ . This choice allows us,, after an appropriate scaling of phase space coordinates, to apply a regular perturbation scheme, which, however, has the consequence that the constructed expansions are not genuine semiclassical expansions, because the correctors depend on the small parameter. The expansions are used in the computation of the wave amplitude for a quartic oscillator with  WKB data of the Gauss-Fresnel type. They both give reasonable  symptomatic approximations at the focal points (caustic) of the oscillator for certain dependence of the coupling constant of the anharmonic potential with the small semiclassical parameter of the Schr\"odinger equation.

\paragraph{Acknowledgements.} EKK has been partially supported by the 
 Research grant 88735, University of Crete (Programme: Graduate fellowships "Heraclitus", funded by the Greek Ministry of Education). GNM has been partially supported by the Archimedes Center for Modeling, Analysis \& Computation (ACMAC), Crete, Greece (grant FP7-REGPDT-2009-1).  GNM would like to thank R. Littlejohn (Berkeley), R. Schubert (Bristol) and A. Athanassoulis (Leicester) for helpful discussions.

 \bigskip
\bigskip
\section*{Appendices}

\medskip
\noindent
\subsection*{Appendix A1:Proof of estimate  (\ref{est1})}
\setcounter{equation}{0}
\renewcommand{\theequation}{A1.\arabic{equation}}

\medskip
For the proof of the estimate  (\ref{est1}) we start from the representation

 \begin{align}
\left[U_{\eps,(xk)}^{-1}\Phi^{\eps}_{nm}\right](\cx,\ck)
&=&\eps
\frac{1}{\pi\eps}\int_{\R}e^{-i\frac{2\ck}{\sqrt{\eps}}\sigma}
\ue_{n}(\sqrt{\eps}\cx+\sigma)\overline{\ue_{m}(\sqrt{\eps}\cx-\sigma)}d\sigma \nonumber \\
&=&\frac{1}{\pi}\int_{\R}e^{-i2\ck
y}\left[U_{\eps}^{-1}\ue_n\right](\cx+y)\left[U_{\eps}^{-1}\ue_m\right](\cx-y)dy \ ,
 \end{align}
and we use the identity
\begin{equation}
\begin{array}{l}
\left[U_{\eps}^{-1}\ue_n\right](\eta)\left[U_{\eps}^{-1}\ue_m\right](\xi)-\sum\limits_{l=0}^{N}\eps^{\frac{l}{2}}
\sum\limits_{\mu=0}^{l}\psi_n^{(\mu)}(\eta)\psi_m^{(l-\mu)}(\xi)\\
=\left(\left[U_{\eps}^{-1}\ue_n\right](\eta)-\sum\limits_{l=0}^{N}\eps^{\frac{l}{2}}\psi_n^{(l)}(\eta)\right)\left(\left[U_{\eps}^{-1}\ue_m\right](\xi)-\sum\limits_{l=0}^{N}\eps^{\frac{l}{2}}\psi_m^{(l)}(\xi)\right)\\
+\left(\left[U_{\eps}^{-1}\ue_n\right](\eta)-\sum\limits_{l=0}^{N}\eps^{\frac{l}{2}}\psi_n^{(l)}(\eta)\right)\sum\limits_{l=0}^{N}\eps^{\frac{l}{2}}\psi_m^{(l)}(\xi)\\
+\left(\left[U_{\eps}^{-1}\ue_m\right](\xi)-\sum\limits_{l=0}^{N}\eps^{\frac{l}{2}}\psi_m^{(l)}(\xi)\right)\sum\limits_{l=0}^{N}\eps^{\frac{l}{2}}\psi_n^{(l)}(\eta)\\
+\sum\limits_{l=N+1}^{2N}\eps^{\frac{l}{2}}\sum\limits_{\mu=l-N}^{N}\psi_n^{(\mu)}(\eta)\psi_m^{(l-\mu)}(\xi) \ .
\end{array}
\end{equation}
Then, we  have
\begin{equation}
\begin{array}{l}
||\left[U_{\eps,(xk)}^{-1}\Phi^{\eps}_{nm}\right]-\sum \limits_{l=0}^{N}\eps^{\frac{l}{2}}\tZnm^{(l)}||^2_{L^2(\R^2)}\\
=\int_{\R}\int_{\R}|\left[U_{\eps,(xk)}^{-1}\Phi^{\eps}_{nm}\right](\cx,\ck)-\sum \limits_{l=0}^{N}\eps^{\frac{l}{2}}\tZnm^{(l)}(\cx,\ck)|^2d\cx d\ck\\
=\int_{\R}\int_{\R}|\frac{1}{\pi}\int_{\R}e^{-i2\ck
y}\left(\left[U_{\eps}^{-1}\ue_n\right](\cx+y)\left[U_{\eps}^{-1}\ue_m\right](\cx-y)-
\sum\limits_{l=0}^{N}\eps^{\frac{l}{2}}\sum\limits_{\mu=0}^{l}\psi_n^{(\mu)}(\cx+y)\psi_m^{(l-\mu)}(\cx-y)\right)dy|^2d\cx d\ck\\
=\int_{\R}\int_{\R}|\left[U_{\eps}^{-1}\ue_n\right](\cx+y)\left[U_{\eps}^{-1}\ue_m\right](\cx-y)-
\sum\limits_{l=0}^{N}\eps^{\frac{l}{2}}\sum\limits_{\mu=0}^{l}\psi_n^{(\mu)}(\cx+y)\psi_m^{(l-\mu)}(\cx-y)|^2d\cx dy\\
=\int_{\R}\int_{\R}|\left[U_{\eps}^{-1}\ue_n\right](\eta)\left[U_{\eps}^{-1}\ue_m\right](\xi)-
\sum\limits_{l=0}^{N}\eps^{\frac{l}{2}}\sum\limits_{\mu=0}^{l}\psi_n^{(\mu)}(\eta)\psi_m^{(l-\mu)}(\xi)|^2d\eta d\xi\\
\le ||[U_\eps^{-1}\ue_n
]-\sum\limits_{l=0}^{N}\eps^{\frac{l}{2}}\psi_n^{(l)}||^2_{L^2(\R)}
||[U_\eps^{-1}\ue_m]-\sum\limits_{l=0}^{N}\eps^{\frac{l}{2}}\psi_m^{(l)}||^2_{L^2(\R)}\\
+||[U_\eps^{-1}\ue_n]-\sum\limits_{l=0}^{N}\eps^{\frac{l}{2}}\psi_n^{(l)}||^2_{L^2(\R)}
\sum\limits_{l=0}^{N}\eps^{\frac{l}{2}}||\psi_m^{(l)}||^2_{L^2(\R)}
+||[U_\eps^{-1}\ue_m]-\sum\limits_{l=0}^{N}\eps^{\frac{l}{2}}\psi_m^{(l)}||^2_{L^2(\R)}
\sum\limits_{l=0}^{N}\eps^{\frac{l}{2}}||\psi_n^{(l)}||^2_{L^2(\R)}\\
+\sum\limits_{l=N+1}^{2N}\eps^{\frac{l}{2}}\sum\limits_{\mu=l-N}^{N}||\psi_n^{(\mu)}||^2_{L^2(\R)}
||\psi_m^{(l-\mu)}||^2_{L^2(\R)} \ .
\end{array}
\end{equation}

Now since $\psi_n^{(\mu)}\in L^2(\R),  \ \mu=0,1,\dots,\
\ n=0,1,\dots$ and
\begin{equation}
||[U_\eps^{-1}\ue_m]-\sum\limits_{l=0}^{N}\eps^{\frac{l}{2}}\psi_m^{(l)}||^2_{L^2(\R)}=O(\eps^{(N+1)/2}) \ ,
\end{equation}
we obtain the desired estimate.

\medskip
\noindent
\subsection*{Appendix A2: Proof of  Theorems 2 \& 3}
\setcounter{equation}{0}
\renewcommand{\theequation}{A2.\arabic{equation}}

\medskip
In the sequel we denote by $\|\cdot \|$ the norms $\|\cdot \|_ {L^2_{r^{\eps}}}$ or $\|\cdot \|_ {L^2}$
and we write them explicitly when the distinction is necessary. The constants  $C_{N}$ are of the generic form $cN!N^{\alpha}$, with  $\alpha>0  \  , \ \ c>0$ (see also  \cite{BR}).

 For the proof of the Theorems \ref{thm1} and \ref{thm2} we need the following lemmas and propositions.

\begin{lemma}\label{faa}(Faa di Bruno formula, \cite{CO} , \cite{BR}, \cite{LERN}).

Let $f:\R^2\rightarrow \R,\ g:\R^2\rightarrow \R^2$  smooth enough functions. For any multi-index   $\nu \in \N^2$ and $z=(\cx,\ck)\in \R^2$ ,
\begin{equation}
\partial^{\nu}\left(f\circ g \right)(z)=\sum_{0\neq |\gamma|\le |\nu|,\ \gamma\in \N^2} \left[ (\partial^{\gamma}f)\circ g \right](z) B_{\nu,\gamma}\left[\partial g\right](z) 
\end{equation}
where
\begin{equation}
B_{\nu,\gamma}\left[\partial g\right]= \nu!\sum_ {\alpha_{\beta}} \prod_{\beta\neq 0} \frac{1}{\alpha_{\beta} !} \left(\frac{\partial^\beta g_1}{\beta!}\right)^{\alpha_{\beta_1}}\left(\frac{\partial^\beta g_2}{\beta!}\right)^{\alpha_{\beta_2}}
\end{equation}
with $\beta=(\beta_1,\beta_2),\ \alpha_{\beta}=(\alpha_{\beta_1}, \alpha_{\beta_2})$
 \[\quad \sum \alpha_{\beta_1}=\gamma_1,\  \sum \alpha_{\beta_2}=\gamma_2,\  \sum \beta_1\alpha_{\beta_1}=\nu_1,\  \sum \beta_2\alpha_{\beta_2}=\nu_2   .\]
\end{lemma}

\medskip
\begin{lemma}\label{hflow}
For the Hamiltonian flow $g_h^t=g_h^t(\cx,\ck)$ of the harmonic oscillator, and for all  $(\cx,\ck)$ and  $t>0$, hold
\begin{enumerate}
\item
\begin{equation}\label{hflow1}
\begin{array}{l}
|\partial_{\ck} g_{h,i}^{\pm t}|\le 1, \quad   |\partial_{\cx} g_{h,i}^{\pm t}|\le 1 \ , \\
\\
 \partial^{\nu}_{\ck} g_{h,i}^{\pm t}=\mathbb{O}_{\nu\times \nu},\quad  \partial^{\nu}_{\cx} g_{h,i}^{\pm t}=\mathbb{O}_{\nu\times \nu},\ \ i=1,2, \   \nu=2,3,\dots \ .
\end{array}
\end{equation}
where $\mathbb{O}_{\nu\times \nu}$ denotes the  ${\nu\times \nu}$ zero matrix.
\item
\begin{equation}\label{hflow2}
|g_{h,i}^{\pm t}(\cx,\ck)|^{\nu}\le( |\cx|+|\ck|)^{\nu},\ i=1,2, \ \nu=1,2,\dots
\end{equation}
\item 
\begin{equation}\label{hflow3}
\left|B_{\nu,\gamma}\left[\partial g_h^{-t}\right]\right|\le  C_{\nu,\gamma}
\end{equation}
 where $B_{\nu,\gamma}$ are defined  in the the previouss Lemma 1, and
 
\item
For all  $f \in \Sh(\R^2)$
\begin{equation} \label{hflow4}
\|f(g_h^{-t})\|=\|f\|
\end{equation}
\end{enumerate}
\end{lemma}
The proof of Lemma 2 is based on a straightforward computation.

\medskip
\begin{lemma}\label{nest}

\begin{enumerate}

\medskip
\item For all $\widetilde{f}_0(\cx,\ck)\in \Sh(\R^2)$ and $m\ge 1$ holds
$$\|(|\cx|+|\ck|)^{m}\partial^{\beta}\widetilde{f}_0\|_{L^2}\le C_{m,\beta}$$

\item For $\widetilde{f}^{\eps}_0(\cx,\ck)$ as in Theorem \ref{thm2} holds

$$\|(|\cx|+|\ck|)^{m}\partial^{\beta}\widetilde{f}^{\eps}_0\|_ {L^2_{r^{\eps}}}\le \eps^{m+1}C_{\beta}$$

\end{enumerate}
where  $\beta=(\beta_1,\beta_2)\in \N^2,\ \ \partial^{\beta}f= \partial^{\beta_1}_{\cx}\partial^{\beta_2}_{\ck} f$
\end{lemma}
\noindent{Proof of Lemma 3:}

\vspace{10pt}
The first part of Lemma 3 is immediate, since $\widetilde{f}_0(\cx,\ck)\in \Sh(\R^2)$. The proof of the second part, relies on direct computation using the explicit form of  $\widetilde{f}^{\eps}_0(\cx,\ck)$. We show the details for the case $\alpha_0(x)=e^{-x^2/2}$ and $S_0(x)=x^2/2$.
 For the case of   $S_0(x)=x$ we proceed similarly.
\noindent
For
$$\widetilde{f}^{\eps}_0(\cx,\ck)=\frac{1}{\pi}e^{-\eps\cx^2}e^{-\frac{(\ck-\cx)^2}{\eps}}$$
and for any $\beta=(\beta_1,\beta_2)\in \N^2$, $\beta\neq 0$, we have
\begin{equation*}
\begin{array}{l}
\partial^{\beta}\widetilde{f}^{\eps}_0(\cx,\ck)=(\sqrt{\eps})^{-|\beta|}\frac{(-2i)^{\beta_2}2^{-|\beta|}e^{i|\beta|\pi/2}}{\sqrt{\pi}}e^{-\eps\cx^2}e^{-\frac{(\ck-\cx)^2}{\eps}}\\
\times \sum_{j=0}^{\beta_1} \left( {\begin{matrix}
   \beta_1  \\
   j  \\
 \end{matrix}}  \right) (-2)^je^{-ij\pi/2} H_j(\sqrt{\eps}\cx)H_{|\beta|-j}\left(\frac{\ck-\cx}{\sqrt{\eps}}\right)
 \end{array}
\end{equation*}
where $H_j$ are the  Hermite polynomials.

The term that dominates  $\partial^{\beta}\widetilde{f}^{\eps}_0(\cx,\ck)$ for small values 
of $\eps$, is
 
$$G_{\beta}^{\eps}(\cx,\ck)=\eps^{-|\beta|}c_{\beta}e^{-\eps\cx^2}e^{-\frac{(\ck-\cx)^2}{\eps}}(\ck-\cx)^{|\beta|} \ ,$$
with $c_{\beta}=(-2i)^{\beta_2}2^{-|\beta|}e^{i|\beta|\pi/2}/\sqrt{\pi}$.

Therefore it is enough to prove a bound for the term $(|\cx|+|\ck|)^{m}G_{\beta}^{\eps}$. We have 

\[\|(|\cx|+|\ck|)^{m}\partial^{\beta}\widetilde{f}^{\eps}_0\|_ {L^2_{r^{\eps}}}\le
\sum_{j=1}^{m} \||\cx|^j|\ck|^{m-j}\partial^{\beta}\widetilde{f}^{\eps}_0\|_ {L^2_{r^{\eps}}}
 \sim\sum_{j=1}^{m}  \||\cx|^j|\ck|^{m-j}G_{\beta}^{\eps} \|_ {L^2_{r^{\eps}}} \ , \] 
and
\begin{equation*}
\begin{array}{l}
\| |\cx|^j|\ck|^{m-j}G_{\beta}^{\eps} \|^2_ {L^2_{r^{\eps}}}\\
= c_{\beta}^2\eps^{-2|\beta|}\int_{\R}\int_{\R}
 |\cx|^{2j}|\ck|^{2(m-j)}(\ck-\cx)^{2|\beta|}e^{-2\eps\cx^2}e^{-\frac{2(\ck-\cx)^2}{\eps}}e^{-\frac{\cx^2+\ck^2 }{\eps^2}}d\cx d\ck\\
=c_{\beta}^2\eps^{-2|\beta|+2|\beta|+2m+2}\int_{\R}\int_{\R}
 |\cx|^{2j}|\ck|^{2(m-j)}(\ck-\cx)^{2|\beta|}e^{-2\eps^3\cx^2}e^{-2\eps(\ck-\cx)^2}
e^{-(\cx^2+\ck^2)}d\cx d\ck\\
=\eps^{2m+2}c_{\beta}^2c^{\eps} \ , \\
\end{array}
 \end{equation*}
where $c^{\eps}\to c < \infty$, as $\eps \to 0$.
This concludes the proof of
 \[\|(|\cx|+|\ck|)^{m}\partial^{\beta}\widetilde{f}^{\eps}_0\|_ {L^2_{r^{\eps}}} \le \eps^{m+1}C_{\beta}\]
\hfill $\blacksquare$

\medskip

\begin{prop}\label{bound1}
\begin{enumerate}
\item For all $f\in \Sh(\R^2)$ and $\nu_j \in \N,\ \ j=1,2\dots$ holds
\begin{equation}\label{Bn}
\|\B_{\nu_1}\left[\B_{\nu_2}\left[\dots\B_{\nu_j}\left[f(g_h^{-t_{j+1}})\right] \dots \right](g_h^{-(t_1-t_{2})}, t_2 )\right](g_h^{-(t-t_1)} ,t_1 ) \|_{L^2}\le C_{\nu_1,\dots,\nu_j}
\end{equation}

\item For $\widetilde{f}^{\eps}_0(\cx,\ck)$ as in Theorem  · \ref{thm2}, and  $\nu_j \in \N,\ \ j=1,2\dots$ holds

\begin{equation}\label{Bn2}
\|\B_{\nu_1}\left[\B_{\nu_2}\left[\dots\B_{\nu_j}\left[f(g_h^{-t_{j+1}})\right] \dots \right](g_h^{-(t_1-t_{2})}, t_2 )\right](g_h^{-(t-t_1)} ,t_1 ) \|_ {L^2_{r^{\eps}}}\le C_{\nu_1,\dots,\nu_j}
\end{equation}
\end{enumerate}
\end{prop}
\noindent{Proof of Proposition 1:}

\vspace{10pt}
Recall that the operators $\B_{\nu_j}$ are given by the formula
$$\B_{\nu_j}= - V^{(\nu_j+2)}(0)\sum\limits_{\lambda_j=0}^{[(\nu_j-1 )/2]+1} c_{\lambda_j,\nu_j} \cx^{\nu_j+1-2\lambda_j}\pklj \ , $$
with
$c_{\lambda_j,\nu_j}=\left(\frac{i}{2}\right)^{2\lambda_j}\frac{1}{(2\lambda_j+1)!(\nu_j+1-2\lambda_j)! } \ .$

By Lemma \ref{hflow}, the left hand side of (\ref{Bn}), (\ref{Bn2}) reads as
$$\| \B_{\nu_1}\left[\B_{\nu_2}\left[\dots\B_{\nu_j}\left[f(g_h^{-t_{j+1}})\right] \dots \right](g_h^{-(t_1-t_{2})}, t_2 ) \right] \parallel \ ,$$
and hence it is enough to estimate this quantity.
First we prove (\ref{Bn}) for all $j$. We give the details only for the cases $j=1$ and $j=2$.
For $j=1$ we have,
 \begin{align*}
\| \B_{\nu_1}\left[f(g_h^{-t_{1}})\right]\|_{L^2}& \le |V^{(\nu_1+2)}(0)| \sum\limits_{\lambda_j=0}^{[(\nu_1-1 )/2]+1}  c_{\lambda_1,\nu_1} \|\cx^{\nu_1+1-2\lambda_1}\pkl f(g_h^{-t_{1}})\|_{L^2} \\&\le \sum\limits_{\lambda_j=0}^{[(\nu_1-1 )/2]+1}  c_{\lambda_1,\nu_1} \sum_{0\neq |\gamma|\le \nu_1}
 C'_{\nu_1,\gamma} \|\cx^{\nu_1+1-2\lambda_1} \left[ (\partial^{\gamma }f)\circ g_h^{-t} \right](z)\left|B_{\nu_1,\gamma}\left[\partial g_h^{-t}\right]\right|\|_{L^2}  \\&
\le \sum\limits_{\lambda_j=0}^{[(\nu_1-1 )/2]+1}  c_{\lambda_1,\nu_1}\sum_{0\neq |\gamma|\le \nu_1}
 C_{\nu_1,\gamma} \| (g_{h,1}^{-t})^{\nu_1+1-2\lambda_1}  \partial^{\gamma }f(z)\| _{L^2}\\&
=\sum\limits_{\lambda_j=0}^{[(\nu_1-1 )/2]+1}  c_{\lambda_1,\nu_1} \sum_{0\neq |\gamma|\le \nu}
 C_{\nu_1,\gamma} \| (|\cx|+|\ck| )^{\nu_1+1-2\lambda_1} \partial^{\gamma }f(z)\|_{L^2}
\end{align*}
where in the first step we used  Faa di Bruno formula, and the then  (\ref{hflow2}),(\ref{hflow3}).

\noindent
The first part of Lemma  \ref{nest} ensures that 
 $$\| (|\cx|+|\ck| )^{\nu_1+1-2\lambda_1} (\partial^{\gamma }f(z)\|_{L^2} \le C_{\lambda_1,\nu_1,\gamma} \ ,$$
 thus
\begin{align*}
\| \B_{\nu_1}\left[f(g_h^{-t_{1}})\right]\|_{L^2}
\le|V^{(\nu_1+2)}(0)| \sum\limits_{\lambda_1=0}^{[(\nu_1-1 )/2]+1} c_{\lambda_1,\nu_1}   C_{\lambda_1,\nu_1} \le C_{\nu_1} \ ,
\end{align*}
which proves   (\ref{Bn}) for $j=1$.

\noindent
For $j=2$, we have

\begin{align*}
 \|& \B_{\nu_1}\left[\B_{\nu_2}\left[f(g_h^{-t_{2}})\right](g_h^{-(t_1-t_{2})})\right]\| \le  \\ & \le |V^{(\nu_1+2)}(0)|  \sum\limits_{\lambda_j=0}^{[(\nu_1-1 )/2]+1} c_{\lambda_1,\nu_1}  \sum_{|\beta^1|\le 2\lambda_1+1} c_{\beta}\|(|\cx|+|\ck|)^{\nu_1+1-2\lambda_1}\partial^{\beta}\left[\B_{\nu_2}\left[f(g_h^{-t_{2}})\right]\right]\|  \\
& \le |V^{(\nu_1+2)}(0)| |V^{(\nu_2+2)}(0)|  \sum\limits_{\lambda_1=0}^{[(\nu_1-1 )/2]+1} \sum\limits_{\lambda_2=0}^{[(\nu_2-1 )/2]+1} c_{\lambda_1,\nu_1}  c_{\lambda_2,\nu_2} \\
&\sum_{|\beta^1|\le 2\lambda_1+1} c_{\beta}\|(|\cx|+|\ck|)^{\nu_1+1-2\lambda_1}\partial^{\beta}\left[ \cx^{ \nu_2+1-2\lambda_2}\partial_{\ck}^{2\lambda_2+1}\left[f(g_h^{-t_{2}})\right]\right]\| \\
&\le\prod_{i=1,2} |V^{(\nu_i+2)}(0)|  \sum\limits_{\lambda_1=0}^{\nu_1^{\ell} }\sum\limits_{\lambda_2=0}^{\nu_2^{\ell}} \sum_{|\beta^1|\le u^1}\sum_{\alpha_1=0}^{\beta_1}\sum_{|\beta^2|\le u^2} C(\nu_1,\nu_2,\lambda_1,\beta^1,\beta^2,\alpha_1)\|(|\cx|+|\ck|)^{\gamma_1}\partial^{\beta^2}f\|
\end{align*}
where
$$\nu_i^{\ell}= [(\nu_i-1 )/2]+1,\ \beta^i\in \N^2,\ u^1=\lambda_1+1,\ u^2=(\beta^1_1-\alpha_1,\beta^1_2+\lambda_2+1) \ ,$$
and
$$ \gamma_1=\nu_1+\nu_2+2-2\lambda_1-2\lambda_2-\alpha_1 \ . $$
Therefore
$$ \| \B_{\nu_1}\left[\B_{\nu_2}\left[f(g_h^{-t_{2}})\right](g_h^{-(t_1-t_{2})})\right]\|_{L^2} \le  C_{\nu_1,\nu_2} \ .$$
 In the same way we can prove it for the general case for all $j\ge 3$, by applying successively  the Faa di Bruno formula, the Leibniz formula and using  Lemma  \ref{hflow},  to get
\begin{align}\label{bb}
&\|\B_{\nu_1}\left[\B_{\nu_2}\left[\dots\B_{\nu_j}\left[f(g_h^{-t_{j+1}})\right] \dots \right](g_h^{-(t_1-t_{2})}, t_2 )\right](g_h^{-(t-t_1)} ,t_1 ) \|_{L^2} \le \\
& \le\prod_{i=1}^{j} |V^{(\nu_i+2)}(0)|  \sum\limits_{\lambda_1=0}^{\nu_1^{\ell} }\sum_{|\beta^1|\le u^1}  \dots \sum\limits_{\lambda_j=0}^{\nu_j^{\ell}}\sum_{\alpha_{j-1} }  \sum_{|\beta^j|\le u^j} C(\nu_i,\lambda_i,\beta^i,\alpha_i)\|(|\cx|+|\ck|)^{\gamma_{j-1}}\partial^{\beta^j}f\|_{L^2}\le  C_{\nu_1,\dots,\nu_j}\nonumber
\end{align}
with
$$\nu_i^{\ell}= [(\nu_i-1 )/2]+1,\ \beta^i\in \N^2\ ,$$
and
$$ \gamma_i=\sum_{k=1}^{i}(\nu_k+1-2\lambda_k)-\sum_{k=1}^{i-1}\alpha_k  \ .$$

For proving the second part we follow the same procedure as before, by using the second part of Lemma \ref{nest}. To proceed we observe that $\|(|\cx|+|\ck|)^{\gamma_{j-1}}\partial^{\beta^j}f\|_ {L^2_{r^{\eps}}}\le \eps^{\gamma_{j-1}+1}C_{\beta^j}$.
This estimate and the fact that   ${\gamma_{j-1}}$ in  (\ref{bb}) is always non-negative, ensure that
\begin{align*}
\|\B_{\nu_1}\left[\B_{\nu_2}\left[\dots\B_{\nu_j}\left[f(g_h^{-t_{j+1}})\right] \dots \right](g_h^{-(t_1-t_{2})}, t_2 )\right](g_h^{-(t-t_1)} ,t_1 ) \|_ {L^2_{r^{\eps}}} \le C_{\nu_1,\dots,\nu_j}  \ , 
\end{align*}
which ends the proof of the proposition. \hfill $\blacksquare$

\vspace{20pt}

\noindent{Proof of Theorem \ref{thm1}:}

 \vspace{10pt}

The $(N+1)$-order remainder (\ref{remthm2}) of  the asymptotic expansion (\ref{expthm2}), that is
\[R^{N+1}(\cx,\ck,t):=\tWe(\cx,\ck,t)-\sum_{l=0}^N \eps^{l/2} {\widetilde Z}^{\eps,(l)}(\cx,\ck,t),\ \ N=0,1,\dots \ ,\]
solves, for aany $N$, the initial value problem
\begin{align}\label{erN}
 \frac{\partial}{\partial t}{R^{N+1}}(\cx,\ck,t)+L_hR^{N+1}(\cx,\ck,t)=-\sum_{\nu=1}^{N+1} \B_{\nu} \left[ R^{N+1-\nu}\right](\cx,\ck,t)\\
 R^{N+1}(\cx,\ck,t)|_{t=0}=0
\end{align}
where $R^{0}=\tWe_h  \ \  \text{and}  \ \   \B_{\nu}:= \B_{\nu}(\cx,\pk)$.

According to Dunhamel's principle the solution of problem (\ref{erN}), is given by the formula
   \[R^{N+1}(\cx,\ck,t)=-\int_0^t \sum_{\nu=1}^{N+1}\eps^{\nu/2}   \B_{\nu}\left[R^{N+1-\nu}\right](g_h^{-(t-s)}(\cx,\ck),s)ds,\ \ N=0,1,\dots\]
Applying the above formula successively  for each $R^{N+1-\nu}$, we have
\begin{align*}
&R^{N+1}(\cx,\ck,t)=\eps^{\frac{N+1}{2}} \left\{   -\int_0^tdt_1 \B_{N+1}\left[\widetilde{f}_0(g_h^{-t_1})\right]\left(g_h^{-(t-t_1)} ,t_1\right)   \right.\\
&+\int_0^tdt_1\int_0^{t_1}dt_2 \B_{N }\left[\B_1\left[\widetilde{f}_0(g_h^{-t_2})\right]\left(g_h^{-(t_1-t_2)},t_2\right)\right]\left(g_h^{-(t-t_1)} ,t_1\right)\\
&+\int_0^tdt_1\int_0^{t_1}dt_2 \B_{N-1 }\left[\B_2\left[\widetilde{f}_0(g_h^{-t_2})\right]\left(g_h^{-(t_1-t_2)},t_2\right)\right]\left(g_h^{-(t-t_1)} ,t_1\right)dt_1dt_2\\
&-\int_0^tdt_1\int_0^{t_1}dt_2\int_0^{t_2}dt_3 \B_{N-1 }\left[\B_1\left[\B_1\left[\widetilde{f}_0(g_h^{-t_3})\right]\left(g_h^{-(t_2-t_3)},t_3\right)\right]\left(g_h^{-(t_1-t_2)},t_2\right)\right]\left(g_h^{-(t-t_1)} ,t_1\right) \\
&+\dots\\
&\left.+(-1)^N  \int_0^tdt_1\int_0^{t_1}dt_2\dots \int_0^{t_N}dt_{N+1}\B_1\left[\B_1\left[\dots \B_1\left[\widetilde{f}_0(g_h^{-t_{N+1}})\right]\dots \right]\left(g_h^{-(t_1-t_2)},t_2\right)\right]\left(g_h^{-(t-t_1)} ,t_1 \right)  \right\}
\end{align*}
where $\widetilde{f}_0(\cx,\ck)$ are the initial data  of the problem (\ref{wh}).

Hence
\begin{align*}
&\|R^{N+1}(\cx,\ck,t)\|_{L^2}\le \eps^{\frac{N+1}{2}}  \left\{  \int_0^tdt_1\|\B_{N+1}\left[f_0(g_h^{-t_1})\right](g_h^{-(t-t_1)} ,t_1)\|_{L^2}  \right. \\
&+\int_0^tdt_1\int_0^{t_1}dt_2\| \B_{N }\left[\B_1\left[f_0(g_h^{-t_2})\right](g_h^{-(t_1-t_2)},t_2)\right](g_h^{-(t-t_1)} ,t_1)\|_{L^2}   \\
&+\dots \\
&+ \left.   \int_0^tdt_1\int_0^{t_1}dt_2\dots \int_0^{t_N}dt_{N+1}\|\B_1\left[\B_1\left[\dots \B_1\left[ f_0(g_h^{-t_{N+1}})\right]\dots \right](g_h^{-(t_1-t_2)},t_2)\right](g_h^{-(t-t_1)} ,t_1 ) \|_{L^2}\right\}
\end{align*}

Since $\widetilde{f}_0(\cx,\ck)\in\Sh(\R^2)$ and $\eps$-independent, the first part of Proposition \ref{bound1} implies that every term  in the right hand side of the above inequality is bounded, and therefore we get
\begin{align*}
\|R^{N+1}(\cx,\ck,t)\|_{L^2}&\le \eps^{\frac{N+1}{2}}  C_{N} \left\{ \int_0^tdt_1 +\int_0^tdt_1\int_0^{t_1}dt_2   +   \dots +\int_0^tdt_1\dots\int_0^{t_N}dt_{N+1}
\right\} \ ,\\
&= \eps^{\frac{N+1}{2}}  C_{N} \sum_{k=1}^{N+1} \frac{t^k}{k!} = \eps^{\frac{N+1}{2}}  C_{N} e^t \ ,
\end{align*} 
which ends the proof.  \hfill $\blacksquare$

\newpage
\noindent
\noindent{Proof of Theorem \ref{thm2}:}

\vspace{10pt}
  
For  $\alpha_0(x)=e^{-x^2/2}$ and   $S_0(x)=x^2/2$, we have
$$
\widetilde{f ^{\eps}}_{0}(\cx,\ck)=\frac{1}{\sqrt{\pi}}e^{-\eps\cx^2}e^{-\frac{(\ck-\cx)^2}{\eps}} \ , 
$$
while for $S_0(x)=x$ we have
$$
\widetilde{f^{\eps}}_{0}(\cx,\ck)=\frac{1}{\sqrt{\pi}}e^{-\eps\cx^2}e^{-\frac{(\ck-1)^2}{\eps}} \ .
$$
In both cases  $\widetilde{f^{\eps}}_0(\cx,\ck) \in \Sh(\R^2)$. So we proceed similarly to the proof  of Theorem 2 and we use the second part of Proposition \ref{bound1}, to  obtain

\begin{align*}
\|R^{N+1}(\cx,\ck,t)\|_ {L^2_{r^{\eps}}}&\le \eps^{\frac{N+1}{2}}  C_{N} \left\{ \int_0^tdt_1 +\int_0^tdt_1\int_0^{t_1}dt_2   +   \dots +\int_0^{t_N}dt_{N+1}
\right\}\Rightarrow\\
&= \eps^{\frac{N+1}{2}}  C_{N} e^t \ ,
\end{align*}
which ends the proof of Theorem \ref{thm2}.
 \hfill $\blacksquare$

\bigskip
\noindent
\section*{Appendix A3: Expansion of $Z^{\eps,(2)}$ }
\setcounter{equation}{0}
\renewcommand{\theequation}{A4.\arabic{equation}}

In this appendix we compute the coefficient   $Z^{\eps,(2)}$ of the harmonic expansion in the case of the quartic oscillator . This  computation is performed by solving
the  problem (\ref{zl}) with $\ell=2$, that is
\begin{eqnarray}\label{ivpz2}
\Bigl(\frac{\partial}{\partial t}+L_h\Bigr){\widetilde Z}^{\eps,(2)}(\cx,\ck,t)=D^{(2)}(\cx,\ck,t)  \nonumber \\
{\widetilde Z}^{\eps,(2)}(\cx,\ck,t)|_{t=0}=0  \ ,
\end{eqnarray}
in two different  ways. 

\paragraph{First way:} We expand ${\widetilde Z}^{\eps,(2)}$ with respect to  the Moyal eigenfunctions $\Psi_nm$ of the harmonic oscillator,
\begin{equation}
{\widetilde Z}^{\eps,(2)}(\cx,\ck,t)=\sum_{n}\sum_{m} z_{nm}^{\eps}(t)\Psi_{nm}(\cx,\ck) \ ,
\end{equation}
and we substitute this series into (\ref{ivpz2}), together with the eigenfunction series (\ref{qlas2}) of   $\tWe_h$, which appears in the $D^{(2)}(\cx,\ck,t)$. 
Then we use the orthogonality of   $\Psi_{nm}$ to derive a hierarchy of  equations for the  coefficients  $z_{nm}^{\eps}(t) \ , \ \ n,m=0,1,\dots \ .$

These equations can be easily integrated because of the polynomial type of the potential  and the special form of Moyal eigenfunctions (Laguerre polynomials), and, after a long and cumbersome computation,  we  get
\begin{align}
 {\widetilde Z}^{\eps,(2)}(\cx,\ck,t)&=-\frac{\mu}{4\pi}Re\left(\sum_{n}\sum_{m} z_{nm}^{\eps}(t)e^{-i(n-m)t}\Psinm(\cx,\ck)\right)+ \nonumber \\
&+\frac{3\mu}{4\pi}tIm\left(\sum_{n}\sum_{m} A_{h,nm,o}^{\eps}(n^2+n)e^{-i(n-m)t}\Psinm(\cx,\ck)\right) \ , \nonumber \\
\end{align}
where
\begin{align*}
z_{nm}^{\eps}(t)&=-\frac{1}{8}(e^{i4t}-1)A_{h,n(m+4),o}^{\eps}((m+1)(m+2)(m+3)(m+4))^{1/2}+\\
&+\frac{1}{8}(e^{-i4t}-1)A_{h,n(m-4),o}^{\eps}(m(m-1)(m-2)(m-3))^{1/2}-\\
&-\frac{1}{4}(e^{i2t}-1)A_{h,n(m+2),o}^{\eps}((4m+6)(m+1)(m+2))^{1/2}+\\
&+\frac{1}{4}(e^{-i2t}-1)A_{h,n(m-2),o}^{\eps}((4m+2)m(m-1))^{1/2} \ ,\\
\end{align*}
with
$$A_{h,nm,o}^{\eps}=(\tWe_0,\Psi_{nm})_{L^2(\R^2_{\cx,\ck})} \ . $$

Then we have
\begin{align*}
\int_{\R}{\widetilde Z}^{\eps,(2)}(\cx,\ck,t)d\ck&=
-\frac{\mu}{4\pi}\sqrt{\eps}\frac{1}{|z^{\eps}| |1-u^{\eps}|}e^{-\cx^2}e^{-2\cx^2Re\left(\frac{u^{\eps}}{1-u^{\eps}}\right)}\times \\
&\times Re\left(-\frac{1}{8}(e^{i4t}-1)(\bar{w}^{\eps})^2g^{\eps}_1(\cx,\ck,t)+\frac{1}{8}(1-e^{i4t})g^{\eps}_2(\cx,\ck,t)-\right.\\
&\left.-\frac{1}{2}(e^{i2t}-1)\bar{w}^{\eps}g^{\eps}_3(\cx,\ck,t)-(e^{-i2t}-1)e^{i2t}g^{\eps}_4(\cx,\ck,t)\right)+\\
&+\frac{3\mu}{4\pi}t\sqrt{\eps}\frac{1}{|z^{\eps}| |1-u^{\eps}|}e^{-\cx^2}e^{-2\cx^2Re\left(\frac{u^{\eps}}{1-u^{\eps}}\right)}
Im\left(g^{\eps}_5(\cx,\ck,t)\right)
\end{align*}
where 
$$u^{\eps}=-w^{\eps}e^{i2t} \ , \ \ \ w^{\eps}=\frac{1-z^{\eps}}{z^{\eps}},\ z^{\eps}=\frac12(1+\eps-i) \ .$$
and
\begin{align*}
g^{\eps}_1(\cx,\ck,t)&=3\frac{(u^{\eps})^2}{(1-u^{\eps})^{2}}+12\cx^2\frac{(u^{\eps})^2}{(1-u^{\eps})^{3}}+4\cx^4\frac{(u^{\eps})^2}{(1-u^{\eps})^{4}}+
6\frac{u^{\eps}}{(1-u^{\eps})}+12\cx^2\frac{u^{\eps}}{(1-u^{\eps})^{2}}+3\\
g^{\eps}_2(\cx,\ck,t)&=\frac{3}{(1-u^{\eps})^{2}}+12\cx^2\frac{1}{(1-u^{\eps})^{3}}+\cx^4\frac{1}{(1-u^{\eps})^{4}}\\
g^{\eps}_3(\cx,\ck,t)&=6\frac{(u^{\eps})^2}{(1-u^{\eps})^{2}}+24\cx^2\frac{(u^{\eps})^2}{(1-u^{\eps})^{3}}+8\cx^4\frac{(u^{\eps})^2}{(1-u^{\eps})^{4}}+
9\frac{u^{\eps}}{(1-u^{\eps})}+18\cx^2\frac{u^{\eps}}{(1-u^{\eps})^{2}}+3\\
g^{\eps}_4(\cx,\ck,t)&=3\frac{u^{\eps}}{(1-u^{\eps})^{2}}+12\cx^2\frac{u^{\eps}}{(1-u^{\eps})^{3}}+4\cx^4\frac{u^{\eps}}{(1-u^{\eps})^{4}}+
\frac{5}{2(1-u^{\eps})}+5\cx^2\frac{1}{(1-u^{\eps})^{2}}\\
g^{\eps}_5(\cx,\ck,t)&=\frac32\frac{(u^{\eps})^2}{(1-u^{\eps})^{2}}+6\cx^2\frac{(u^{\eps})^2}{(1-u^{\eps})^{3}}+2\cx^4\frac{(u^{\eps})^2}{(1-u^{\eps})^{4}}+\frac{3u^{\eps}}{(1-u^{\eps})}+3\cx^2\frac{u^{\eps}}{(1-u^{\eps})} \ .
\end{align*}

At the focal points
 $(x_{\nu},t_{\nu})=(0,\nu\pi-\pi/4),\ \nu=1,2,...$, 
$$|1-u^{\eps}_{\nu}|=|1-i\frac{1-\bar{z}^{\eps}}{\bar{z}^{\eps}}|=\frac{1}{|z^{\eps}|}\frac{\eps}{\sqrt{2}} \ ,$$
and thus we get
\begin{equation}
\int_{\R}{\widetilde Z}^{\eps,(2)}(0,\ck,t_{\nu})d\ck=\frac{\sqrt{2}}{\pi\eps^{3/2}}\mu (\beta+\beta^{\eps}) \ ,
\end{equation}
with
$$\beta^{\eps}=\left( -\eps^2(\frac{3\pi}{16}( \mu-1/4)+3)+\eps\frac{17}{2}\right) \ \  \text{and}\ \
\beta=\frac{\pi}{8} ( \mu-1/4)-3 \ .$$

\medskip
\noindent
\paragraph{Second way:} By  Dunhamel's principle, the solution of (\ref{ivpz2}) is given by
\begin{equation}
{\widetilde Z}^{\eps,(2)}(\cx,\ck,t)=\int_{0}^{t}D^{(2)} (q(\cx,\ck,t-s ),p ( \cx,\ck,t-s ),s )ds \ ,
\end{equation}
where
\begin{equation}
D^{(2)}(\cx,\ck,s)=-\mu\left[\frac14\cx \pkkk -\cx^3 \pk\right]\We_h(\cx,\ck,s)=-\B_2(\cx,\pk){\widetilde W}^{\eps}_{h}(\cx,\ck,t) \ . 
\end{equation}

With the aid of symbolic computations with MAPLE, for any  $(\cx,\ck,t)$ we obtained the expression 
\begin{align}
{\widetilde Z}^{\eps,(2)}(\cx,\ck,t)&=\mu\We_h(\cx,\ck,t)\times \nonumber \\
&\times\left[\frac{1}{\eps^3}F_1(\cx,\ck,t)+\frac{1}{\eps^2}F_2(\cx,\ck,t)+\frac{1}{\eps}F_3(\cx,\ck,t)+\right. \nonumber\\
&+\left. F_4(\cx,\ck,t)+\eps F_5(\cx,\ck,t)+ \eps^2 F_6(\cx,\ck,t)+\eps^3F_7(\cx,\ck,t)\right] \ ,  \nonumber \\
\end{align}
where
\begin{align*}
F_1(\cx,\ck,t)=&2[\ck c(t)+\cx s(t)]^3(\cx f_1(t)+\ck f_2(t))\\
F_2(\cx,\ck,t)=&-8[\ck c(t)+\cx s(t)]f_3(t)\\
F_3(\cx,\ck,t)=&-2[\ck c(t)+\cx s(t)]\left(\ck^3f_4(t)+3\cx\ck^2f_5(t)+3\cx^2\ck f_6(t)+\cx^3f_7(t)+\right.\\
&\left.+3[\ck c(t)+\cx s(t)][\cx\cos(t)-\ck\sin(t)](\ck f_8(t)+\cx f_9(t))\right)\\
\end{align*}
and 
\begin{align*}
F_4(\cx,\ck,t)=&8\left((\cx\cos(t)-\ck\sin(t))f_{10}(t)+[\ck c(t)+\cx s(t)]f_{11}(t)\right)\\
F_5(\cx,\ck,t)=&2(\cx\cos(t)-\ck\sin(t)) 3[\ck c(t)+\cx s(t)](\cx f_{12}(t)+\ck f_{13}(t))+\\
&+2(\cx\cos(t)-\ck\sin(t))(\cx^3f_{14}(t)+3\cx^2\ck f_{15}(t)+3\cx \ck^2f_{16}(t)+\ck^3 f_{17}(t)))\\
F_6(\cx,\ck,t)=&-8(\cx\cos(t)-\ck\sin(t))f_{18}(t)\\
F_7(\cx,\ck,t)=&-2(\cx\cos(t)-\ck\sin(t))^3(\cx f_{19}(t)+\ck f_{20}(t))\\
\end{align*}
with $c(t)=\cos(t)+\sin(t),\ s(t)=\sin(t)-\cos(t)$, and $f_j(t)$ being also nonlinear combinations of harmonic  functions of time $t$.

Returning  to  the  variables $(x,k)$, we obtain

\begin{align}\label{expz2}
Z^{\eps,(2)}(x,k,t)&=\mu\We_H(x,k,t)\times\left[\frac{1}{\eps^5}F_1(x,k,t)+\frac{1}{\eps^{5/2}}F_2(x,k,t)+\frac{1}{\eps^3}F_3(x,k,t)\right. \nonumber \\
&\left.+\frac{1}{\eps^{1/2}}
F_4(x,k,t) +\frac{1}{\eps}F_5(x,k,t)+\eps^{3/2}F_6(x,k,t)+\eps F_7(x,k,t)\right] \ . \nonumber \\
\end{align}

The integration of the expansion  (\ref{expz2}), which is a rather long and complicated computation, leads to the same result.

\end{document}